\documentclass[largeformat]{interact}

\suppldatatrue

\usepackage[T1]{fontenc}
\usepackage{lmodern}
\usepackage{natbib}
\bibpunct[, ]{(}{)}{;}{a}{}{,}

\usepackage{amsmath}
\usepackage{graphicx}
\usepackage{booktabs}
\usepackage{siunitx}
\usepackage{array}
\usepackage[hidelinks]{hyperref}

\newcommand{\figpaneltag}[1]{(#1)}

\newlength{\figurebeforesep}
\newlength{\figurecaptionsep}
\newlength{\figureaftersep}
\newlength{\figurepanelgap}
\newlength{\setuppanelheight}
\newlength{\setupdiagramheight}
\begin{document}
\renewcommand{\topfraction}{0.90}
\renewcommand{\bottomfraction}{0.80}
\renewcommand{\textfraction}{0.05}
\renewcommand{\floatpagefraction}{0.85}
\renewcommand{\dbltopfraction}{0.90}
\renewcommand{\dblfloatpagefraction}{0.85}

\articletype{Original Article}

\title{HoloCMA: A Holographic Eye on Coarse-Mode Aerosols}

\author{
\name{Nikil Krishnakumar\textsuperscript{a,b},
Ryne A. Juidici\textsuperscript{b},
Nicholas Bravo-Frank\textsuperscript{a},
Shibo Wang\textsuperscript{b},
Qisheng Ou\textsuperscript{b},
Francisco J. Romay\textsuperscript{b},
Wing Lai\textsuperscript{a},
Chongai Kuang\textsuperscript{c},
Naruki Hiranuma\textsuperscript{d},
David Y.H. Pui\textsuperscript{b}, and
Jiarong Hong\textsuperscript{a,b}\thanks{CONTACT Jiarong Hong. Email: jhong@umn.edu}}
\affil{\textsuperscript{a}Particle4X, Inc., 219 SE Main Street, Suite 202, Minneapolis, MN 55414, USA; \\
\textsuperscript{b}Department of Mechanical Engineering, University of Minnesota, 111 Church Street SE, Minneapolis, MN 55455, USA; \\
\textsuperscript{c}Environmental Science and Technologies Department, Brookhaven National Laboratory, Bldg. 815E, Upton, NY 11973-5000, USA; \\
\textsuperscript{d}Department of Physics, The University of Texas at El Paso, 500 W University, El Paso, TX 79968, USA}
}

\maketitle
\thispagestyle{empty}
\pagestyle{empty}

\begin{abstract}
Coarse-mode aerosols (CMAs), including pollen, spores, and dust, remain difficult to characterize in situ because conventional instruments do not jointly resolve particle geometry and number concentration. We present HoloCMA, which combines digital inline holography, computational reconstruction, and deep learning-assisted analysis for particle-resolved measurements. Its optical configuration is designed for particles from approximately \SI{5.0}{\micro\meter} to the millimeter scale. Active sampling operates at up to \SI{30.0}{L\per\minute}, and the active-sampling module can be detached for open-path operation. With an NVIDIA GeForce RTX 5070 Laptop GPU (\SI{8}{GB} VRAM), HoloCMA reports equivalent circular diameter (ECD) and number concentration in real time without sustained queue buildup up to \SI{11.4}{particles\per\centi\meter\cubed}. Additional computing capacity could extend real-time operation to the workflow limit of \SI{100.0}{particles\per\centi\meter\cubed}. We evaluated HoloCMA with monodisperse polystyrene latex spheres, sodium chloride and ammonium sulfate crystals, and oleic acid droplets spanning nominal diameters of \SIrange{4.9}{11.8}{\micro\meter}, using an Aerodynamic Particle Sizer (APS) for comparison. Across 12 comparisons, absolute differences between HoloCMA and corrected APS geometric-mean diameters ranged from less than 0.1 to \SI{1.0}{\micro\meter}, with mean absolute and mean absolute relative differences of \SI{0.5}{\micro\meter} and 8.4\%. HoloCMA derives geometric size directly from reconstructed contours, whereas converting APS aerodynamic diameter requires material-specific inputs. The APS-to-HoloCMA concentration ratio generally decreased with particle size, consistent with stronger size-dependent transport and counting losses in the APS, without definitive attribution. HoloCMA therefore provides continuous CMA measurements that combine geometric size, concentration, and retained particle images, supporting classification and long-duration atmospheric, environmental, and indoor-air monitoring.

\end{abstract}

\begin{keywords}
coarse-mode aerosol; digital inline holography; particle-resolved imaging; geometric particle sizing; aerosol number concentration; deep learning; real-time aerosol monitoring
\end{keywords}

\twocolumn

\section{Introduction}
\label{sec:intro}

Coarse-mode aerosols (CMAs) are atmospheric particles with diameters spanning the micrometer and supermicrometer size ranges and include mineral dust, sea salt, pollen, fungal spores, and volcanic ash~\citep{seinfeld2016atmospheric}. Although generally less numerous than submicrometer aerosols, CMAs are abundant in the atmosphere, can dominate aerosol mass, and affect climate by scattering and absorbing solar radiation, acting as giant cloud condensation nuclei or ice-nucleating particles, and modifying cloud microphysics, precipitation, and radiative forcing~\citep{myhre2013anthropogenic,kok2017dust,boucher2013clouds}. Their importance extends beyond climate. Pollen, fungal spores, and mineral dust contribute to respiratory and occupational exposure~\citep{damato2010effects,douwes2003bioaerosol}. Wildfire smoke is also linked to adverse respiratory outcomes~\citep{reid2016wildfire,aguilera2025wildfirepm}. Airborne spores and pollen are relevant to crop disease monitoring and management~\citep{mccartney1994spores,vanderheyden2021monitoring}. Particles near and above \SI{5.0}{\micro\meter} are also important in cleanroom classification and pharmaceutical contamination control, where personnel and garments are major emission sources~\citep{iso14644cleanrooms,meng2024pharmaceutical}. Accurate real-time characterization of individual CMAs is therefore relevant across fixed atmospheric observatories, such as the Atmospheric Radiation Measurement (ARM) program~\citep{mather2013arm} and AmeriFlux/FLUXNET sites~\citep{baldocchi2001fluxnet}, as well as mobile platforms including uncrewed aerial systems and ship-based stations.

Despite this broad importance, real-time in situ characterization of individual CMAs remains technically challenging, particularly when size, morphology, and concentration must be resolved simultaneously. Existing techniques each address only part of this multidimensional measurement need. Filter-based gravimetric analysis provides aerosol mass integrated over a sampling time interval but does not resolve individual particle size, morphology, or short-timescale variability~\citep{mcmurry2000review}. The Aerodynamic Particle Sizer (APS, Model 3321) provides real-time aerodynamic size distributions over part of the coarse mode, with a nominal sizing range of \SIrange{0.5}{20.0}{\micro\meter}. However, relating aerodynamic diameter to geometric diameter requires assumptions about particle density and dynamic shape factor, and transport and counting efficiencies decrease for larger particles~\citep{baron2011aerosol,tsi_aps3321,volckens2005counting,vasilatou2023extending}. Optical Particle Counters (OPCs) provide number concentration and equivalent optical diameter, but the reported size depends on particle refractive index and shape. Consequently, spherical-particle calibration can misrepresent the geometric dimensions of nonspherical particles~\citep{Hinds2022Aerosol}. Single-particle fluorescence instruments add useful biological aerosol discrimination, including fluorescence-based identification of individual pollen grains, although fluorescence thresholds and interference from nonbiological aerosols complicate classification~\citep{kiselev2013pollen,savage2017wibs}. Taken together, these limitations leave a specific instrumentation gap: simultaneous, continuous in situ measurement of individual CMAs with geometric size, morphology, and number concentration resolved across an extended size range.

Digital inline holography (DIH) offers a promising approach for addressing this measurement need by enabling high-throughput in situ imaging of individual particles over an extended depth of field~\citep{katz2010applications,fugal2009cloud,berg2022tutorial,sauvageat2020pollen}. Using a coherent light source and a digital camera, DIH records interference patterns, or holograms, formed when light scattered by a particle interferes with the unscattered portion of the reference beam. Numerical reconstruction brings particles at different axial positions into focus from a single hologram, providing particle number, three-dimensional position, geometric size, shape, and orientation over an extended depth of field~\citep{katz2010applications,berg2022tutorial}. This individual-particle imaging avoids the density and dynamic-shape-factor conversion required to infer geometric diameter from aerodynamic diameter and the refractive-index dependence of equivalent optical sizing. Reconstructed holograms also retain diffraction- and phase-related signatures that are sensitive to particle refractive index, providing optical information beyond geometric size~\citep{berg2022tutorial,beres2024microplastics}. When combined with machine learning, morphological and optical information encoded in holographic signatures can also support automated particle detection, segmentation, and classification~\citep{sauvageat2020pollen,shyamkumar2025review}. Important challenges nevertheless remain when DIH is translated from laboratory particle imaging to field-ready CMA monitoring. Such an instrument must combine a sufficiently large sampling volume, robust aerosol handling, automated reconstruction, concentration retrieval, and real-time processing. One existing instrument that addresses this need is the Swisens Poleno, a field-deployable DIH system developed for automated pollen monitoring. It combines holographic imaging with a virtual impactor to enhance coarse-particle sampling~\citep{sauvageat2020pollen,lieberherr2021assessment}. However, the concentration factor is strongly size dependent. Reference-chamber experiments measured factors of 7.85 at \SI{5.0}{\micro\meter} and 123.79 at \SI{10.0}{\micro\meter}, substantially below the theoretical maximum of approximately 1000~\citep{lieberherr2021assessment}. This size dependence complicates quantitative concentration retrieval across broad particle-size distributions. More broadly, conventional reconstruction and focus evaluation across many candidate depths remain computationally intensive, which further complicates real-time analysis of large hologram datasets~\citep{shyamkumar2025review}.

To address this gap, we present the HoloCMA, a holographic instrument for real-time in situ characterization of individual CMAs. The HoloCMA combines DIH with an automated data-processing pipeline and is designed to characterize airborne particles from approximately \SI{5.0}{\micro\meter} to millimeter-scale sizes. The present processing limit is \SI{100.0}{particles\per\centi\meter\cubed}. The system reports equivalent circular diameter (ECD) and number concentration while retaining reconstructed particle images for morphology analysis. We validate its geometric-sizing capability using polystyrene latex (PSL) spheres, sodium chloride (\(\mathrm{NaCl}\)) crystals, ammonium sulfate (\((\mathrm{NH}_4)_2\mathrm{SO}_4\)) crystals, and oleic acid (\(\mathrm{C}_{18}\mathrm{H}_{34}\mathrm{O}_2\)) droplets spanning \SIrange{4.9}{11.8}{\micro\meter}, with measurements from the APS converted to a geometric basis for comparison. We also compare number concentrations from the HoloCMA and the APS, documenting a size-dependent trend between the two instruments. Section~\ref{sec:methods} describes the HoloCMA system and data-processing pipeline. Section~\ref{sec:validation} presents the validation experiments and particle-comparison methods. Section~\ref{sec:results} reports the sizing and concentration results. Section~\ref{sec:conclusion} provides the conclusion and discussion.

\section{System Description}
\label{sec:methods}

\subsection{Hardware}

\begin{figure}[htbp]
\centering
\includegraphics[width=0.96\columnwidth]{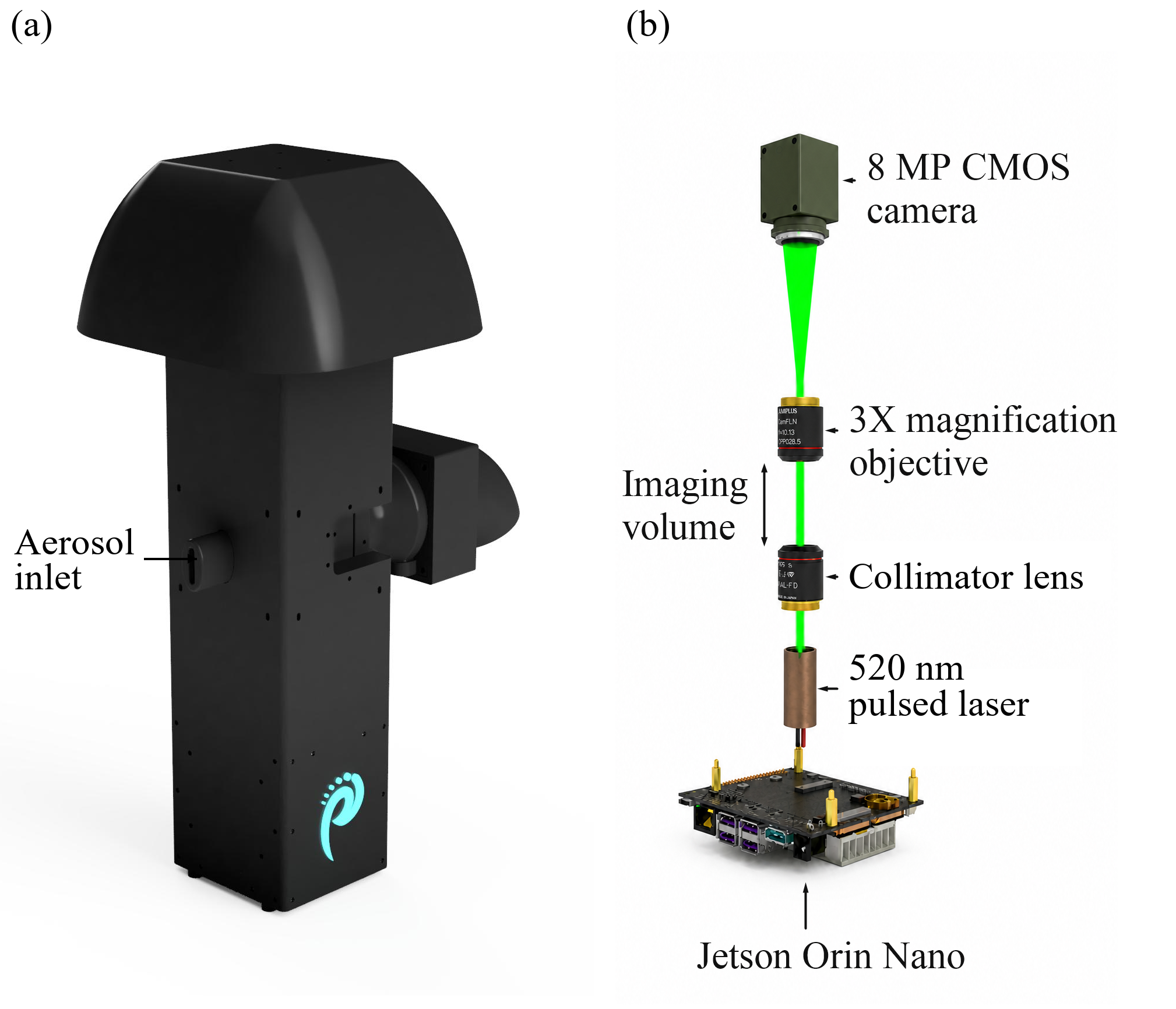}
\caption{Overview of the HoloCMA instrument. (a) Rendered view of the hardware assembly with the aerosol inlet indicated. (b) Exploded view of the principal imaging and computing components with the optical imaging volume indicated.}
\label{fig:holocma_overview}
\end{figure}
The HoloCMA is a compact instrument for field deployment built around a digital inline holography (DIH) imaging unit and an onboard data processing unit (Fig.~\ref{fig:holocma_overview}). The optical path begins with a \SI{520}{nm}, \SI{50}{mW} pulsed laser diode. The emitted light is conditioned by a collimator lens and then passes through a \(3\times\) magnification objective before reaching an 8-megapixel (MP) CMOS camera, which records the resulting holograms. This configuration achieves a raw spatial sampling of \SI{0.9}{\micro\meter\per pixel} and a frame rate of \SI{20.0}{Hz}. The optical beam illuminates a sample volume of \(2.7 \times 2.7 \times 12\) \si{mm^3}, and the optical configuration is designed to accommodate individual particles from approximately \SI{5.0}{\micro\meter} to the millimeter scale. The present processing workflow is designed for operating concentrations up to \SI{100.0}{particles\per\centi\meter\cubed}. The instrument supports active sampling with controlled flow and can be converted to open-path operation without forced flow by detaching the active-sampling module. The standard active operating flow rate is \SI{20.0}{L\per\minute}, and the system can operate at active flow rates up to \SI{30.0}{L\per\minute}. Holograms recorded by the camera are passed to an NVIDIA Jetson Orin Nano operating in \SI{25}{W} mode with \SI{8}{GB} LPDDR5 memory and \SI{1}{TB} internal SSD storage, which handles onboard acquisition, background enhancement, particle detection, and region of interest (ROI) extraction. Particle ROIs are automatically transferred (via Wi-Fi or Ethernet) to an integrated laptop equipped with an NVIDIA GeForce RTX 5070 Laptop GPU (\SI{8}{GB} VRAM), which performs particle focusing, particle sizing, and quality control. This laptop sustains real-time computation without queue buildup at concentrations up to \SI{11.4}{particles\per\centi\meter\cubed}. Additional computing capacity can extend real-time processing to the current operating limit of \SI{100.0}{particles\per\centi\meter\cubed}.

\begin{figure*}[!t]
\centering
\includegraphics[width=0.96\textwidth]{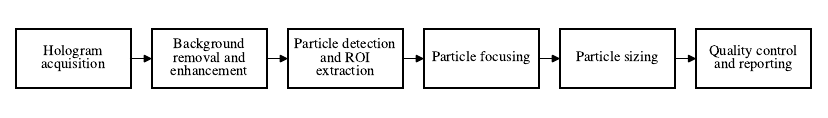}
\caption{Overview of the processing workflow implemented in the HoloCMA. Following hologram acquisition and background removal, particle detection and ROI extraction use a heatmap-based detector to identify particle candidates. Particle focusing uses a kurtosis criterion to select the focused reconstruction, particle sizing determines ECD from the segmented contour, and quality control removes artifacts before reporting size distributions and number concentration.}
\label{fig:processing_workflow}
\end{figure*}

\subsection{Data Processing Pipeline}

Raw holograms captured by the camera are processed through a multistage workflow that converts each acquired frame into particle-level size and concentration records (Fig.~\ref{fig:processing_workflow}). The workflow follows the general structure used in recent DIH deep learning pipelines, where hologram enhancement, particle detection, focusing, and segmentation are combined to extract particle-level information from raw holographic measurements~\citep{bravo2024ham,bravofrank2024bacteria}. The stages below define how the HoloCMA converts raw holograms into detected particle ROIs, focused particle images, and contour-based size measurements.

\textit{Hologram acquisition.} The camera records raw inline holograms at \SI{20.0}{Hz} and transfers each frame to the onboard processing unit. The optical acquisition geometry and imaged sample volume are described in Section~\ref{sec:methods}.

\textit{Background removal and enhancement.} Each acquired frame undergoes moving window background subtraction to suppress static optical artifacts, nonuniform illumination, and persistent signatures from particles deposited on the imaging window~\citep{bravo2024ham,bravofrank2024bacteria}. By reducing stationary signatures from the imaging-window glass in the raw frame, the enhanced image emphasizes newly sampled airborne particles and preserves their holographic signatures for downstream analysis.

\textit{Particle detection and ROI extraction.} Particle candidates are identified using a convolutional heatmap detector applied to the enhanced hologram. The detector uses a MobileNetV3-Small backbone to extract multiscale feature maps at \(1/4\), \(1/8\), \(1/16\), and \(1/32\) of the image resolution~\citep{howard2019mobilenetv3}. A lightweight feature-pyramid decoder then combines these maps through \(1\times1\) lateral convolutions and top-down feature fusion before a single-channel sigmoid heatmap head produces the spatial particle-likelihood map~\citep{lin2017fpn}. The resulting heatmap identifies particle locations for ROI extraction. For real-time edge operation, the trained network is deployed through a TensorRT engine on the Jetson Orin Nano~\citep{nvidia2023tensorrt}, with a PyTorch model retained as a fallback~\citep{paszke2019pytorch}. Extracted particle ROIs and associated metadata are automatically transferred during acquisition to the integrated laptop (via Wi-Fi or Ethernet), which performs the remaining analysis steps as particle detections are produced.

\textit{Particle focusing.} Each particle ROI is numerically reconstructed at candidate planes across the imaging depth using band-limited angular-spectrum propagation~\citep{matsushima2009bandlimited}. The HoloCMA evaluates each candidate plane using an excess-kurtosis focus metric~\citep{sulvaran2026benchmarking}. The plane with the highest central-region kurtosis is identified as the in-focus plane, and the ROI is numerically reconstructed at this plane for particle characterization.

\textit{Particle sizing.} After particle focusing determines the particle depth, the corresponding ROI is reconstructed from the raw hologram. The image intensity is normalized to the local background, and the particle boundary is extracted from the focused intensity distribution using contour-based segmentation~\citep{suzuki1985border}. The same contour-area-to-ECD calculation is applied to every particle and reported as its geometric size.

\textit{Quality control and reporting.} A final filtering stage rejects particles whose focused images fail to meet minimum sharpness, size, or contour quality thresholds. This stage also suppresses non-aerosol artifacts that can persist after enhancement, including repeated stationary detections from imaging-window deposits and bright-core laser or optical flicker events that do not have the dark focused signature expected for particles. Accepted detections are archived and combined into particle-size distributions and number-concentration records.

Particle classification is not implemented in the present workflow. However, the retained focused images preserve morphology-, diffraction-, and phase-related optical features that could support future particle differentiation and classification~\citep{sauvageat2020pollen,shyamkumar2025review}.

\begin{table*}[!t]
\caption{\centering Manufacturer-certified PSL and FMAG geometric reference diameters and their roles in HoloCMA validation.}
\label{tab:validation_particles}
\centering
\small
\setlength{\tabcolsep}{8pt}
\renewcommand{\arraystretch}{1.08}
\begin{tabular}{@{}>{\raggedright\arraybackslash}p{0.18\textwidth}>{\raggedright\arraybackslash}p{0.22\textwidth}p{0.50\textwidth}@{}}
\toprule
Particle type & Reference diameter(s) (\si{\micro\meter}) & Validation role \\
\midrule
PSL spheres & 7.0 & Manufacturer-certified monodisperse spherical sizing reference with a well-defined diameter and density, used to benchmark the sizing accuracy of the HoloCMA. \\
\cmidrule(lr){1-3}
Sodium chloride crystals & 3.0, 4.9, 6.1, 7.0, 9.7 & Crystalline salt commonly generated with the FMAG and used to evaluate geometric sizing of irregular crystals across the tested diameter range. \\
\cmidrule(lr){1-3}
Ammonium sulfate crystals & 4.9, 6.1, 7.7 & Crystalline salt with a composition distinct from sodium chloride, used to test whether the sizing approach remains consistent for a second crystalline aerosol. \\
\cmidrule(lr){1-3}
Oleic acid droplets & 4.9, 6.9, 9.9, 11.8 & Nearly spherical liquid droplets used to evaluate the sizing response of the HoloCMA for liquid particles and provide a contrast with the crystalline aerosols. \\
\bottomrule
\end{tabular}
\end{table*}

\section{Validation Experiments}
\label{sec:validation}

\subsection{Overview}

Laboratory validation of the HoloCMA was conducted at the Particle Technology Laboratory at the University of Minnesota. Particle generation used a flow-focusing monodisperse aerosol generator (FMAG, Model 1520). Four aerosol types were selected for the experiments: PSL spheres, sodium chloride crystals, ammonium sulfate crystals, and oleic acid droplets. Together, these materials span crystalline particles, a spherical polymer sizing standard, and liquid droplets over a nominal size range of \SIrange{3.0}{11.8}{\micro\meter}. The experimental setup generated each particle condition and directed the aerosol simultaneously to the HoloCMA and a TSI Aerodynamic Particle Sizer (APS, Model 3321). The subsequent particle comparison accounted for the different diameter definitions reported by the two instruments. The APS reports aerodynamic diameter, whereas the HoloCMA reports an image-derived geometric diameter. We therefore converted the APS measurements to a geometric basis using material-specific effective density, dynamic shape factor, and liquid-droplet corrections as applicable. These corrections enable a direct instrument-to-instrument comparison between the HoloCMA and the APS.

\subsection{Particle Samples}

The validation included four particle types compatible with FMAG generation: PSL spheres as a polymer sizing standard, sodium chloride crystals, ammonium sulfate crystals, and oleic acid droplets. Table~\ref{tab:validation_particles} summarizes the manufacturer-certified PSL-sphere diameter, the FMAG geometric reference diameters for the generated crystals and droplets, and the validation role of each particle type. The corresponding FMAG generation parameters and nominal-diameter uncertainties are provided in Supplementary Note S1.

The PSL-sphere reference used for the nominal \SI{7.0}{\micro\meter} condition consisted of NIST-traceable PSL spheres supplied in water (Thermo Scientific Duke Standards 3K/4K particle counter standard, catalog 4K-07, lot 204390, Fremont, CA, USA). The manufacturer-certified particle diameter was \(7.0 \pm 0.1\)~\si{\micro\meter}, with a coefficient of variation of 1.0\%. This narrow, traceable size distribution provided the reference population used to evaluate sizing by the HoloCMA.

Sodium chloride crystals were selected as a commonly generated crystalline salt for evaluating the sizing of irregularly shaped crystals by the HoloCMA across FMAG geometric reference diameters of 3.0, 4.9, 6.1, 7.0, and \SI{9.7}{\micro\meter}. The \SI{3.0}{\micro\meter} sodium chloride crystal condition specifically probed the lower detection boundary of the HoloCMA. Although the HoloCMA detected some sodium chloride crystals during this experiment, the measured response did not meet expectations for reliable detection at \SI{3.0}{\micro\meter}, supporting a practical optical detection threshold of approximately \SI{5.0}{\micro\meter}.

Ammonium sulfate crystals provided a second crystalline salt with a composition distinct from sodium chloride crystals. The FMAG geometric reference diameters of 4.9, 6.1, and \SI{7.7}{\micro\meter} were used to examine whether the sizing approach of the HoloCMA remained consistent for ammonium sulfate crystals, including a \SI{4.9}{\micro\meter} ammonium sulfate crystal condition near the instrument's lower detection threshold.

Oleic acid droplets were selected to evaluate the sizing of liquid droplets by the HoloCMA at FMAG geometric reference diameters of 4.9, 6.9, 9.9, and \SI{11.8}{\micro\meter}, providing a contrast with the crystalline aerosols. Oleic acid droplets are expected to be nearly spherical, giving a dynamic shape factor of unity and an effective density equal to the bulk-liquid density. After accounting for instrument-specific droplet effects, their geometric and aerodynamic diameters are expected to be similar because the density of oleic acid droplets is close to the aerodynamic reference density~\citep{Hinds2022Aerosol,baron2008droplets}. Oleic acid droplets were evaluated after the other particle types, as droplet deposits on the APS focusing nozzle can alter the measured sizes of subsequently tested aerosols~\citep{baron2008droplets}. For all generated populations, an APS-measured geometric standard deviation below 1.2 was targeted. Only the \SI{9.7}{\micro\meter} sodium chloride crystal distribution exceeded this criterion. The FMAG was cleaned and flushed before each new particle size or type was introduced.

\begin{figure*}[!t]
\centering
\begin{minipage}[t]{0.51\textwidth}
\vspace{0pt}
\noindent\makebox[\linewidth][l]{\figpaneltag{a}}\par
\vspace{\figurepanelgap}
\centering
\includegraphics[height=\setupdiagramheight,width=\linewidth,keepaspectratio]{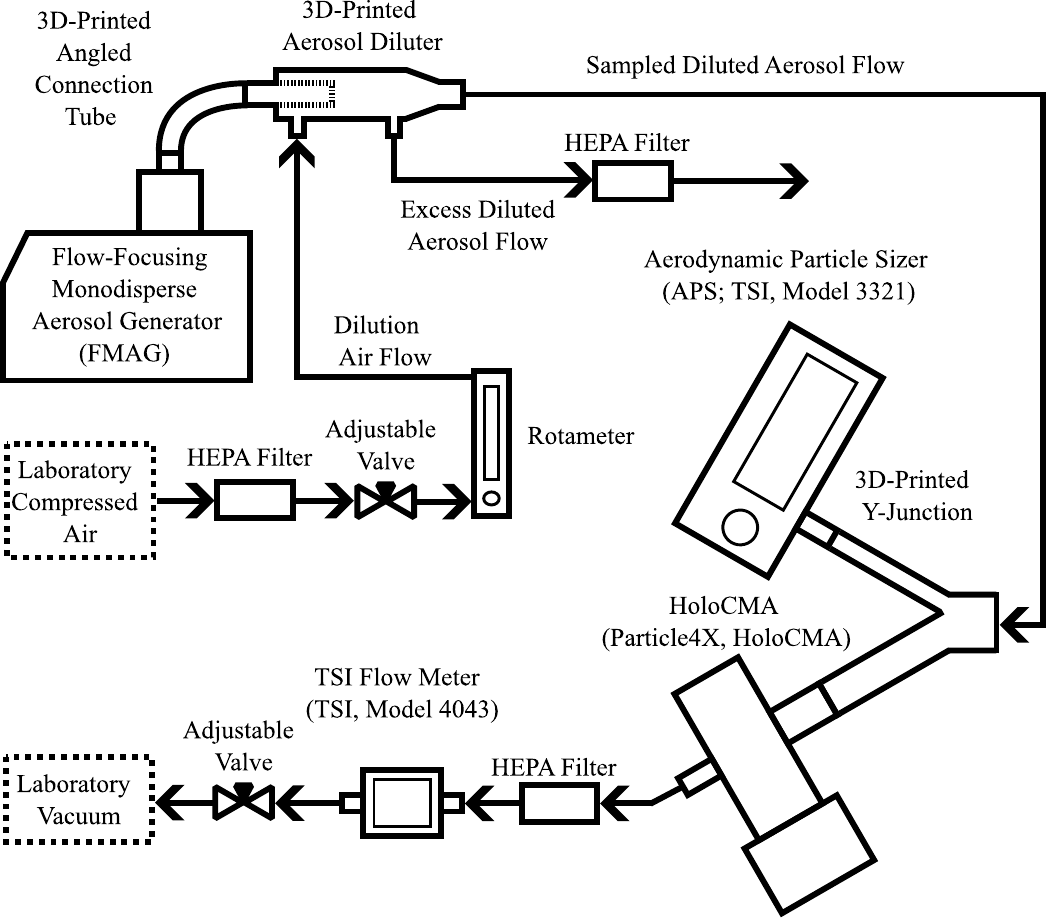}
\end{minipage}
\hfill
\begin{minipage}[t]{0.47\textwidth}
\vspace{0pt}
\noindent\makebox[\linewidth][l]{\figpaneltag{b}}\par
\vspace{\figurepanelgap}
\vspace{0.23\textwidth}
\centering
\includegraphics[height=\setuppanelheight,width=\linewidth,keepaspectratio]{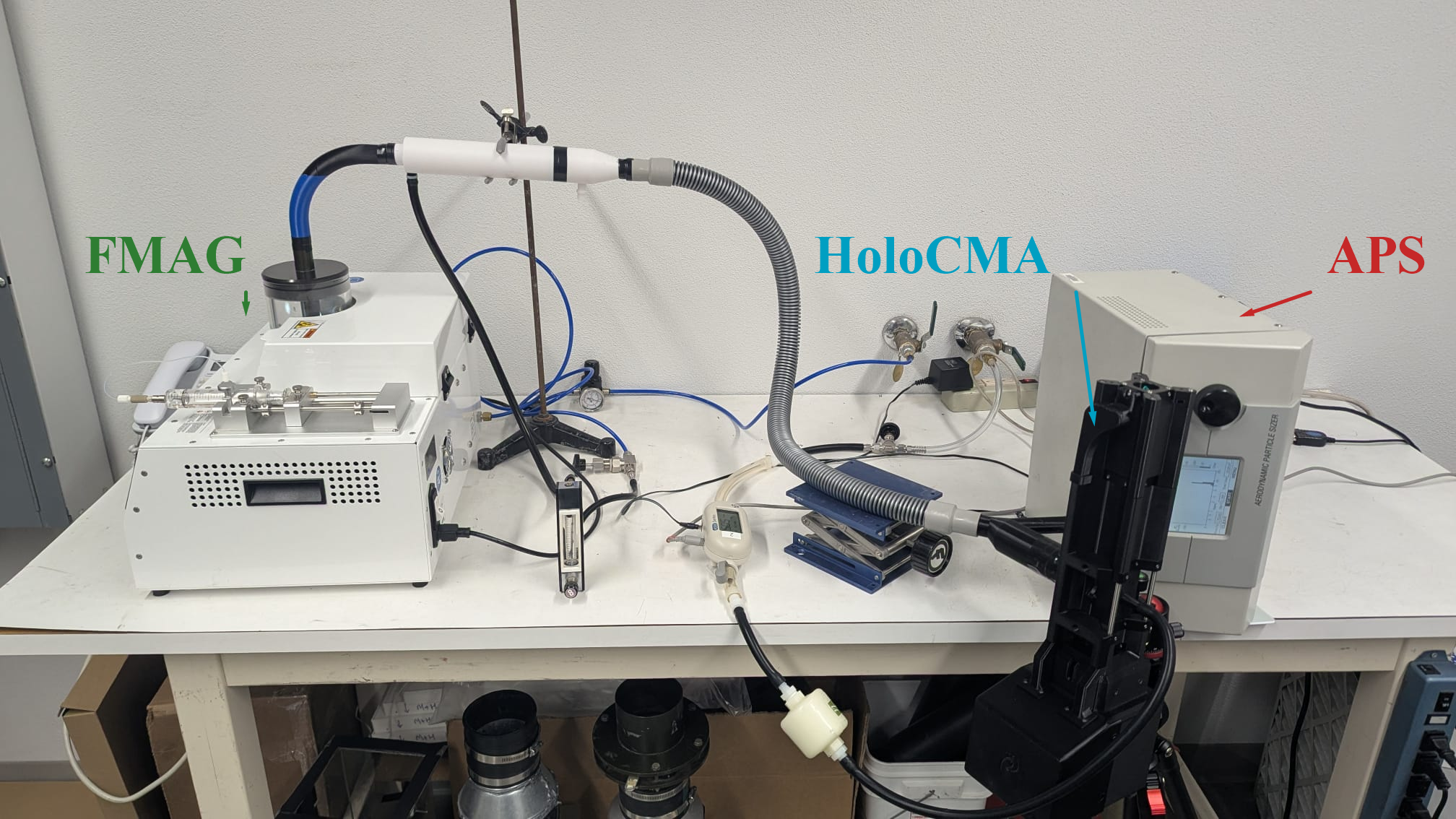}
\end{minipage}

\caption{Experimental setup used for validation of the HoloCMA against the APS. (a) Schematic of aerosol generation, dilution, flow control, and parallel sampling by the HoloCMA and the APS. (b) Laboratory implementation of the aerosol generation and sampling arrangement.}
\label{fig:experimental_setup}
\end{figure*}

\subsection{Experimental Setup}

Test aerosols were introduced using the FMAG, as shown schematically in Fig.~\ref{fig:experimental_setup}. Particle solutions or suspensions were prepared at the appropriate concentrations and loaded into the FMAG. The nominal droplet diameter generated by the FMAG is determined by~\citep{duan2016generation}
\begin{equation}
  d_d = \left(\frac{6Q}{\pi f}\right)^{1/3},
\end{equation}
where $d_d$ is the nominal droplet diameter, $Q$ is the liquid flow rate controlled by the FMAG syringe pump, and $f$ is the droplet generation frequency controlled by the FMAG. The FMAG liquid-flow verification is summarized in Supplementary Note S2. By selecting a satisfactory droplet diameter and using a known solution concentration of the chosen particle type, the nominal geometric particle diameter is given by
\begin{equation}
    d_p = C^{1/3}d_d,
\end{equation}
where $d_p$ is the nominal geometric particle diameter and $C$ is the solution concentration. In this manner, the FMAG was tuned to the desired geometric particle diameter prior to each experimental run.

Directly downstream of the FMAG was a 3D-printed curved sampling line to turn the vertical FMAG outlet flow into a horizontal flow path. Immediately after this, a 3D-printed diluter was used. The 3D-printed diluter allowed filtered air to be combined with aerosol-laden air, diluting the overall number concentration. To avoid overpressurizing the APS and the HoloCMA, an outlet downstream of the dilution air inlet removed excess air from the system. The diluted sample aerosol flow was then transported through a large-diameter hose to a custom 3D-printed Y-junction, with one branch directed to the APS inlet and the other to the HoloCMA inlet. Because the inlet flow rates of the HoloCMA and the APS were \SI{25.0}{L\per\minute} and \SI{5.0}{L\per\minute}, respectively, the Y-junction used different branch dimensions to reduce velocity mismatch and support approximately isokinetic sampling between the two instruments.

For these validation experiments, the HoloCMA was operated in active-sampling mode at an inlet flow rate of \SI{25.0}{L\per\minute}, rather than its standard active operating flow rate of \SI{20.0}{L\per\minute}. Open-path operation was not evaluated because the instrument-to-instrument validation required the HoloCMA and the actively aspirated APS to sample simultaneously from the same controlled aerosol stream. The experimental flow was maintained using laboratory vacuum, monitored with an inline flow meter (TSI, Model 4043), and remained within the supported active operating range up to \SI{30.0}{L\per\minute}. This inlet flow rate should be distinguished from the volumetric sampling rate of the HoloCMA. The HoloCMA images a \(2.7 \times 2.7 \times 12\) \si{mm^3} optical volume at \SI{20.0}{Hz}, corresponding to an analyzed sample volume of approximately 6.3~L/hour. For the APS, the total inlet flow rate of \SI{5.0}{L\per\minute} was comprised of a \SI{1.0}{L\per\minute} aerosol sample flow and a \SI{4.0}{L\per\minute} sheath flow maintained by the instrument's internal pumps and controls~\citep{tsi_aps3321}. The APS inlet and sample flow rates were verified at the start of each evaluation period using a bubble flow meter.

\begin{figure*}[!t]
\centering
\includegraphics[width=0.96\textwidth]{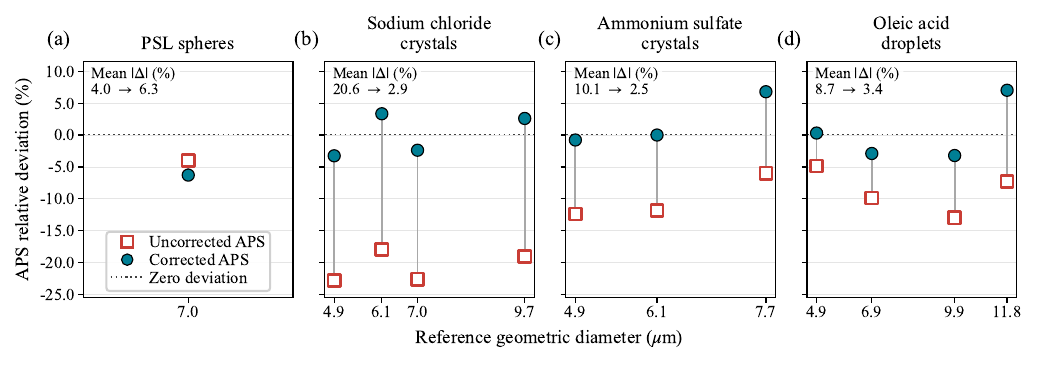}
\caption{Relative deviation of the APS before and after aerodynamic-to-geometric diameter correction for (a) PSL spheres, (b) sodium chloride crystals, (c) ammonium sulfate crystals, and (d) oleic acid droplets. Signed relative deviation is defined as $100(d_{\mathrm{APS}}-d_{\mathrm{ref}})/d_{\mathrm{ref}}$, where $d_{\mathrm{ref}}$ is the manufacturer-certified diameter for PSL spheres and the FMAG geometric reference diameter for the generated crystals and droplets. Negative values indicate underestimation. Gray lines connect the paired uncorrected aerodynamic and corrected geometric values at each reference size. Panel annotations report the mean absolute relative deviation before and after correction.}
\label{fig:aps_correction}
\end{figure*}

During validation, the HoloCMA recorded raw hologram sequences locally without concurrent processing. The stored images were processed offline using a consistent analysis configuration. Sampling intervals ranged from 3 to 8 minutes per trial, and each reported run concentration represents the time-averaged concentration over the complete acquisition interval rather than an instantaneous transient value. All sizing and concentration plots include the three runs collected with each instrument for each condition. After each trial, the HoloCMA imaging window was cleaned to remove deposited particles and prevent carryover between trials. All sizing and concentration data reported here are based on simultaneous sampling by the APS and the HoloCMA.

\subsection{Particle Comparison Methods}
\label{sec:particle_comparison}

For each accepted detection from the HoloCMA, the particle boundary area, $A$, was obtained from the selected contour and converted to ECD,
\begin{equation}
d_{\mathrm{ECD}}=2\sqrt{\frac{A}{\pi}}.
\end{equation}
Pixel dimensions were converted using the calibrated spatial sampling of \SI{0.9}{\micro\meter\per pixel}. Across the validation datasets, 1.4\% of the focused particle images could not be sized. For a run containing $N_f$ analyzed frames, the accumulated optical sample volume was calculated as
\begin{equation}
V_{\mathrm{tot}}=N_fV_{\mathrm{frame}}, \quad V_{\mathrm{frame}}=(2.7\times2.7\times12)~\si{mm^3}.
\end{equation}
The run-averaged number concentration from the HoloCMA was then calculated as
\begin{equation}
n_{\mathrm{run}}=\frac{N_{\mathrm{accepted}}}{V_{\mathrm{tot}}},
\end{equation}
where $N_{\mathrm{accepted}}$ is the number of particle detections retained after quality control. The pulsed illumination effectively freezes particle motion during each acquisition, allowing every hologram to be treated as an instantaneous sample of the optical volume. This volume-based calculation produces one concentration averaged over the complete acquisition interval of the HoloCMA rather than a frame-by-frame time series. For the present comparison, the APS was configured to report the concentration averaged over the corresponding run. At \SI{20.0}{Hz}, the accumulated imaged volume corresponds to approximately 6.3~L/hour. Because the HoloCMA concentrations are calculated from the imaged optical volume rather than the active inlet flow rate, inlet-flow variation does not directly rescale the retrieved concentration, although it can affect particle transport and deposition. Distributions from the HoloCMA were grouped into \SI{0.9}{\micro\meter} bins, consistent with the object-plane spatial sampling, while data from the APS retained the instrument's native logarithmically spaced channels. For each run, the area-normalized number-size density, $\tilde{n}(D_p)$, was calculated by dividing the number concentration in bin $i$, $n_i$, by its bin width, $\Delta D_{p,i}$, and then normalizing by the total number concentration:
\begin{equation}
\tilde{n}(D_{p,i})=\frac{n_i/\Delta D_{p,i}}{\sum_j n_j},
\qquad
\sum_i \tilde{n}(D_{p,i})\Delta D_{p,i}=1.
\end{equation}
The ratio $n_i/\Delta D_{p,i}$ is the bin-width-corrected $dn/dD_p$, and $\tilde{n}(D_{p,i})$ is its area-normalized form in \si{\per\micro\meter}. This procedure accounts for the nonuniform APS channel widths and is applied consistently to both instruments. Accordingly, the width-weighted integrated area is unity, although individual density values may exceed \SI{1.0}{\per\micro\meter}. For all analysis, the first APS bin was neglected due to non-physical noise observed across trials. Since the evaluation is concerned with \SI{3.0}{\micro\meter} and larger particles, neglecting this bin has no influence on the presented results or subsequent discussion. For both instruments, the geometric mean was calculated from channel midpoints, $d_i$, and concentration weights, $n_i$, following standard aerosol size-distribution analysis~\citep{baron2011aerosol,Hinds2022Aerosol}, as
\begin{equation}
d_g=\exp\left(\frac{\sum_i n_i\ln d_i}{\sum_i n_i}\right).
\end{equation}
For the three-material comparisons, $\tilde{n}(D_p)$ was calculated independently for each run. The size-distribution profiles show the mean of the three area-normalized runs, and the error bars show one standard deviation across runs within each size bin for both instruments.

The PSL-sphere aerosolization produced a separate small-particle population in both instruments. Consequently, the PSL-sphere size-distribution, geometric-mean, and concentration comparisons were restricted to the matched \SIrange{5.0}{9.0}{\micro\meter} interval containing the nominal \SI{7.0}{\micro\meter} spheres. All reported PSL-sphere distributions, geometric means, concentrations, and uncertainty estimates use this interval for both instruments.

Replicated concentrations are reported as the three-run mean and sample standard deviation for both instruments at every condition. For each paired run $r$, the concentration ratio and its three-run mean were calculated as
\begin{equation*}
R_{n,r}=\frac{n_{\mathrm{APS},r}}{n_{\mathrm{HoloCMA},r}},
\qquad
R_n=\frac{1}{3}\sum_{r=1}^{3}R_{n,r}.
\end{equation*}
Figure~\ref{fig:concentration_comparison}d reports $R_n$ and its sample standard deviation across the three paired runs. Values above unity indicate a higher concentration from the APS, while values below unity indicate a higher concentration from the HoloCMA.

The different inlet flow rates of the HoloCMA and the APS required different Y-junction branch diameters, which produced different gravitational deposition losses upstream of the instrument inlets. To place the concentration measurements on a common upstream basis, the APS concentrations were adjusted using the relative branch penetrations predicted by the gravitational deposition model of \citet{pich1972theory}. The model was applied using the measured flow rates and branch geometry under the assumptions of circular, laminar sampling lines and gravitational settling as the dominant branch-loss mechanism. Supplementary Note S3 provides the model equations, branch dimensions, and Reynolds numbers. This adjustment accounts for predicted differential transport through the two sampling branches but does not correct either instrument's internal counting efficiency.

Because the HoloCMA reports ECD and the APS reports aerodynamic diameter, the APS channel diameters were converted to equivalent-volume geometric diameters before comparison. The APS time-of-flight measurement is calibrated against unit-density PSL spheres, so conversion for particles of different density or shape requires the particle effective density and dynamic shape factor~\citep{Hinds2022Aerosol,duan2016generation}. For FMAG-generated sodium chloride crystals and ammonium sulfate crystals, the effective density after spray drying can differ from the bulk crystalline density because of internal voids or non-ideal crystal packing~\citep{tiwari2017fmag}. The dynamic shape factor accounts for the additional aerodynamic drag produced by non-spherical geometry. This correction is particularly relevant to spray-dried sodium chloride crystals, which can form cubic crystals and agglomerates~\citep{zelenyuk2006agglomerates}.

We selected effective-density and dynamic-shape-factor values from previously published measurements of sodium chloride crystals and ammonium sulfate crystals~\citep{tiwari2017fmag,Kuwata2009,zelenyuk2006agglomerates}. \citet{zelenyuk2006agglomerates} reported experimentally derived values for near-micrometer particles. Although this range is smaller than that evaluated here, both quantities appear to approach asymptotic values that allow estimates to be extended to the present size range. We used an effective density of 1.60~g~cm$^{-3}$ and a dynamic shape factor of 1.15 for sodium chloride crystals, and an effective density of 1.55~g~cm$^{-3}$ and a dynamic shape factor of 1.08 for ammonium sulfate crystals. These values are similar to those reported by \citet{Kuwata2009}. The APS aerodynamic diameter, $d_a$, was converted to equivalent-volume geometric diameter, $d_g$, using
\begin{equation}
d_{g} = d_{a}\sqrt{\frac{\chi\rho_0}{\rho_{\mathrm{eff}}}},
\end{equation}
where $\rho_0$ is the aerodynamic reference density, $\rho_{\mathrm{eff}}$ is the particle effective density, and $\chi$ is the dynamic shape factor~\citep{Hinds2022Aerosol,tiwari2017fmag,duan2016generation}. Slip correction was neglected because of the large particle sizes and ambient-pressure conditions.

For oleic acid droplets, the equivalent-volume conversion used a dynamic shape factor of unity and an effective density equal to the bulk-liquid density of approximately 0.9~g~cm$^{-3}$, consistent with spherical droplets. We also applied an empirical correction to the APS aerodynamic diameter before conversion to equivalent-volume diameter. Acceleration through the APS focusing nozzle can deform oleic acid droplets into oblate spheroids, increasing their aerodynamic drag and shifting the measured time of flight toward smaller diameters~\citep{baron2008droplets}. Oleic acid droplet deposition on the focusing nozzle can further increase the local velocity and shift the measured size downward~\citep{baron2008droplets}. We therefore applied the empirical size-shift correction of \citet{baron2008droplets} to recover the corrected aerodynamic diameter before converting it to equivalent-volume diameter. Supplementary Note S4 describes its application.

Figure~\ref{fig:aps_correction} summarizes the effect of these corrections relative to the manufacturer-certified diameter for PSL spheres and the FMAG geometric reference diameters for sodium chloride crystals, ammonium sulfate crystals, and oleic acid droplets. For PSL spheres, density conversion shifted the APS mean from approximately \SI{6.7}{\micro\meter} aerodynamic diameter to \SI{6.6}{\micro\meter} geometric diameter, changing the absolute relative deviation from the manufacturer-certified diameter from 4.0\% to 6.3\%. For sodium chloride crystals, ammonium sulfate crystals, and oleic acid droplets, the corrections reduced the mean absolute relative deviations of the APS from 20.6\%, 10.1\%, and 8.7\% to 2.9\%, 2.5\%, and 3.4\%, respectively. All subsequent comparisons use the corrected equivalent-volume diameters from the APS.

Together, these procedures place the measurements from the HoloCMA and the APS on a common geometric-diameter basis and define consistent calculations for particle size and concentration. The following section applies this comparison framework first to the traceable PSL-sphere reference and then to the generated sodium chloride crystals, ammonium sulfate crystals, and oleic acid droplets.

\section{Results}
\label{sec:results}

\begin{figure}[htbp]
\centering
\begin{minipage}{0.98\columnwidth}
\centering
\includegraphics[width=0.96\columnwidth]{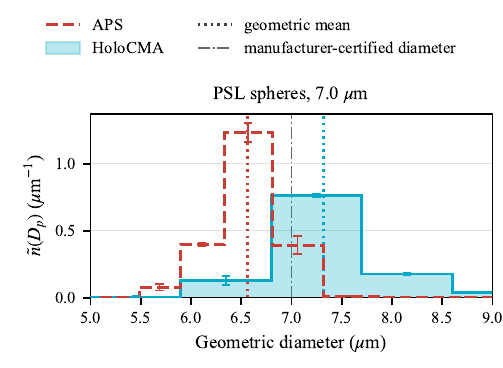}
\end{minipage}
\caption{Area-normalized PSL-sphere number-size densities, $\tilde{n}(D_p)$, for the nominal \SI{7.0}{\micro\meter} condition over the matched \SIrange{5.0}{9.0}{\micro\meter} interval. Data from the APS retain the native logarithmically spaced channels, while data from the HoloCMA use \SI{0.9}{\micro\meter} bins. For each run, concentration was divided by bin width to obtain $dn/dD_p$ before area normalization. Accordingly, the width-weighted integrated area equals unity, not the sum of the displayed bin heights. Size-distribution profiles show the mean of three normalized runs, error bars indicate one standard deviation, dotted lines indicate geometric means of \SI{6.6}{\micro\meter} for the APS and \SI{7.3}{\micro\meter} for the HoloCMA, and the gray dash-dot line indicates the manufacturer-certified \SI{7.0}{\micro\meter} diameter.}
\label{fig:psl_distribution}
\end{figure}

Following the corrections described in Section~\ref{sec:particle_comparison} and summarized in Fig.~\ref{fig:aps_correction}, the HoloCMA and APS measurements were compared on a common geometric-diameter basis. We first evaluated the HoloCMA using the NIST-traceable \SI{7.0}{\micro\meter} PSL spheres, which provide the most direct sizing benchmark because of their manufacturer-certified diameter and narrow size distribution. Figure~\ref{fig:psl_distribution} compares the area-normalized number-size distributions, $\tilde{n}(D_p)$, measured simultaneously by the two instruments over the matched \SIrange{5.0}{9.0}{\micro\meter} interval. The APS distribution retains its native logarithmically spaced channels, whereas the HoloCMA measurements are grouped into \SI{0.9}{\micro\meter} bins.

Both instruments clearly resolved the primary PSL-sphere population near the manufacturer-certified diameter. The modal estimates from the raw particle-size data were \SI{6.9}{\micro\meter} for the APS and \SI{7.1}{\micro\meter} for the HoloCMA, placing both within approximately 1.5\% of the manufacturer-certified \SI{7.0}{\micro\meter} diameter. The geometric means were \SI{6.6}{\micro\meter} for the APS and \SI{7.3}{\micro\meter} for the HoloCMA, corresponding to deviations of 6.3\% below and 4.6\% above the manufacturer-certified diameter, respectively. The relative difference between the two geometric means was 11.6\%.

The close agreement between the distribution modes indicates that both instruments located the dominant PSL-sphere population accurately. The difference between the geometric means may partly arise from the discrete size channels of the APS and from sampling variability in the sparsely populated tails of the HoloCMA distribution, which contained fewer particles. Thus, the PSL-sphere comparison establishes agreement for the principal particle population while also illustrating how binning and particle count can influence distribution-derived statistics. This initial benchmark provides a reference point for testing the method across particles with different compositions and physical properties.

\begin{figure*}[!t]
\centering
\includegraphics[width=0.96\textwidth]{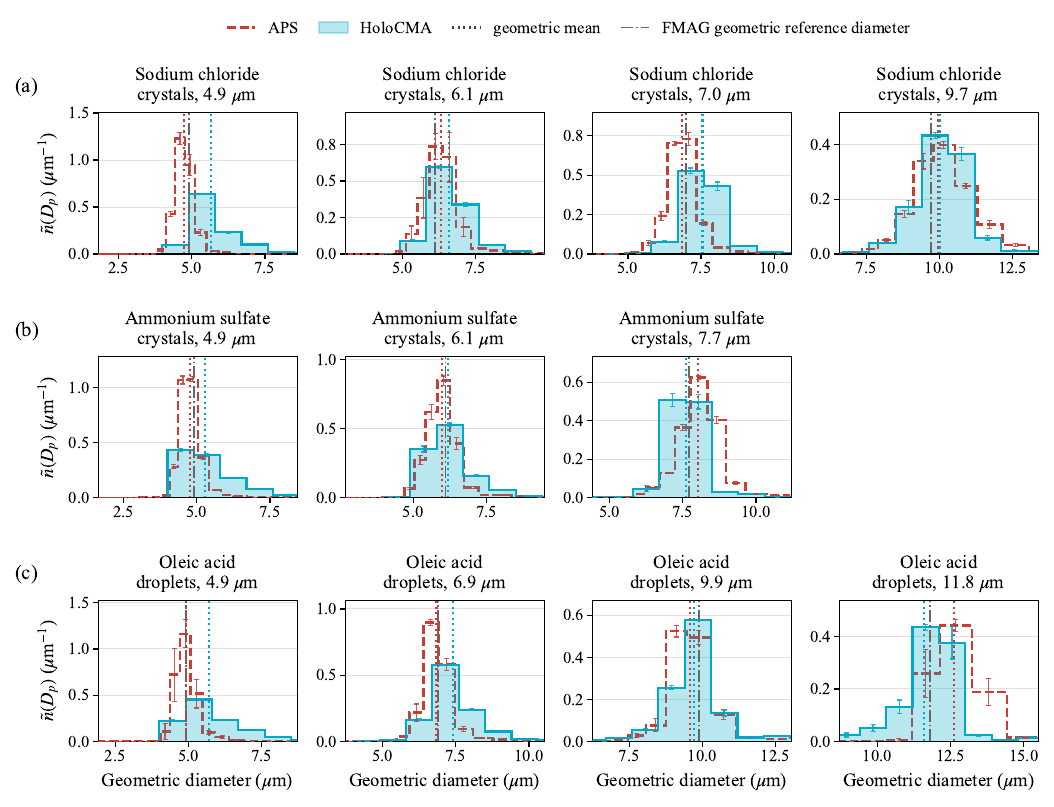}
\caption{Size distribution comparison between the APS and the HoloCMA for sodium chloride crystals, ammonium sulfate crystals, and oleic acid droplets. Distributions from the APS retain their corrected native channels, while distributions from the HoloCMA use \SI{0.9}{\micro\meter} ECD bins. For each run, concentration was divided by bin width to obtain $dn/dD_p$ and then area-normalized as $\tilde{n}(D_p)$. The width-weighted integrated area equals unity, while individual density values may exceed \SI{1.0}{\per\micro\meter}. Size-distribution profiles show the mean of three normalized runs, error bars indicate one standard deviation, dotted lines indicate the geometric mean diameter, and gray dash-dot lines indicate the FMAG geometric reference diameters shown in the panel titles.}
\label{fig:main_size_distributions}
\end{figure*}

Figure~\ref{fig:main_size_distributions} broadens the size-distribution comparison to sodium chloride crystals, ammonium sulfate crystals, and oleic acid droplets over FMAG geometric reference diameters from \SI{4.9}{\micro\meter} to \SI{11.8}{\micro\meter}. Each panel presents the mean area-normalized number-size distribution, $\tilde{n}(D_p)$, from three runs, together with the geometric means reported by the two instruments and the FMAG geometric reference diameter. Across all three material series, the central regions of the HoloCMA and corrected APS distributions generally overlapped. Both instruments also captured the progressive shift toward larger diameters as the FMAG geometric reference diameter increased. These consistent shifts demonstrate that the HoloCMA tracked changes in particle size across sodium chloride crystals, ammonium sulfate crystals, and oleic acid droplets.

To quantify the agreement without relying only on absolute differences, the absolute relative difference was calculated as
\[
100\frac{|d_{g,\mathrm{HoloCMA}}-d_{g,\mathrm{APS}}|}{d_{g,\mathrm{APS}}},
\]
where $d_{g,\mathrm{HoloCMA}}$ and $d_{g,\mathrm{APS}}$ are the geometric mean diameters reported by the HoloCMA and the corrected APS, respectively.
Including the PSL-sphere condition, the absolute geometric-mean differences across the 12 comparisons ranged from less than \SI{0.1}{\micro\meter} to \SI{1.0}{\micro\meter}, with a mean absolute difference of \SI{0.5}{\micro\meter} and a mean absolute relative difference of 8.4\%. Ten of the 12 conditions differed by approximately 12\% or less. For sodium chloride crystals, the relative differences ranged from 0.7\% to 19.5\%. Those for ammonium sulfate crystals ranged from 3.4\% to 11.1\%, while those for oleic acid droplets ranged from 1.3\% to 16.2\%. For oleic acid droplets, the absolute differences ranged from \SI{0.1}{\micro\meter} at \SI{9.9}{\micro\meter} to \SI{1.0}{\micro\meter} at \SI{11.8}{\micro\meter}. Ten of the 12 absolute differences were within the \SI{0.9}{\micro\meter} HoloCMA histogram-bin width, although the bin width is not a measurement-uncertainty bound. The largest differences occurred near the lower detection boundary at \SI{4.9}{\micro\meter} for sodium chloride crystals and oleic acid droplets. No consistent monotonic relationship was observed between the inter-instrument difference and particle size.

The FMAG geometric reference diameters provide an additional basis for assessing the size-dependent trends. Agreement between the HoloCMA geometric mean and the FMAG geometric reference diameter generally improved as particle size increased. For sodium chloride crystals, the relative deviation decreased from 15.7\% at \SI{4.9}{\micro\meter} to 3.3\% at \SI{9.7}{\micro\meter}. For ammonium sulfate crystals, it decreased from 8.1\% at \SI{4.9}{\micro\meter} to 1.3\% at \SI{7.7}{\micro\meter}. For oleic acid droplets, it decreased from 16.5\% at \SI{4.9}{\micro\meter} to 1.8\% at \SI{11.8}{\micro\meter}. This trend is consistent with the increasing detectability and more robust contour estimation of larger particle images.

\begin{figure*}[!t]
\centering
\includegraphics[width=0.96\textwidth]{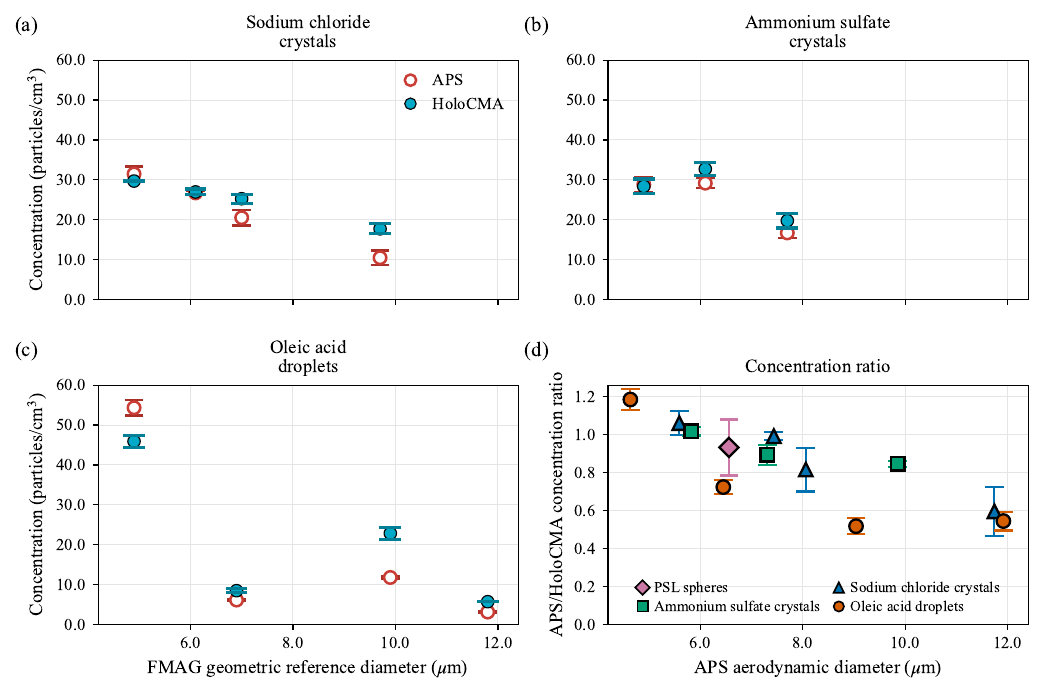}
\caption{Concentration comparison between the APS and the HoloCMA. Panels (a) to (c) show three-run mean concentrations for sodium chloride crystals, ammonium sulfate crystals, and oleic acid droplets. Concentrations from the APS include the differential branch-loss adjustment described in Particle Comparison Methods. Error bars show one standard deviation. Panel (d) shows the mean paired-run concentration ratio, \(R_n\), with error bars across three paired runs. Ratios above unity indicate a higher concentration from the APS, while ratios below unity indicate a higher concentration from the HoloCMA. The PSL-sphere concentrations were calculated over the matched \SIrange{5.0}{9.0}{\micro\meter} interval for both instruments. The PSL-sphere condition is identified by its manufacturer-certified diameter, while the generated-particle conditions are identified by their FMAG geometric reference diameters. Markers in panel (d) are positioned using the geometric mean aerodynamic diameter from the APS.}
\label{fig:concentration_comparison}
\end{figure*}

These comparisons should nevertheless be interpreted with several qualifications. The FMAG geometric reference diameters represent nominal generation conditions rather than independently measured particle diameters. Internal voids and nonideal crystal packing can increase the physical dimensions of salt particles relative to values inferred from their nominal generation conditions, as discussed in Supplementary Note S5. In addition, conversion of the APS aerodynamic diameter to a geometric diameter depends on the assumed effective density, dynamic shape factor, and, for oleic acid droplets, the applied droplet correction. Residual differences may therefore reflect uncertainties in both the APS conversion and the focus and contour estimation used by the HoloCMA. The smaller error bars observed for the HoloCMA in some panels may also result from its coarser \SI{0.9}{\micro\meter} bins and should not be interpreted as evidence of greater repeatability. These sizing comparisons provide the context needed to interpret the concentration results in Fig.~\ref{fig:concentration_comparison}.

Figure~\ref{fig:concentration_comparison} summarizes the three-run mean number concentrations measured by the HoloCMA and the branch-loss-adjusted APS. Panels (a) through (c) show the concentrations for sodium chloride crystals, ammonium sulfate crystals, and oleic acid droplets. Figure~\ref{fig:concentration_comparison}d presents the paired-run APS-to-HoloCMA concentration ratio, $R_n$, as a function of the APS geometric-mean aerodynamic diameter. For oleic acid droplets, these APS diameters include the liquid-droplet correction described in Section~\ref{sec:particle_comparison}.

The closest agreement occurred near \SI{5.0}{\micro\meter}. At the \SI{4.9}{\micro\meter} conditions, $R_n$ was 1.1 for sodium chloride crystals, 1.0 for ammonium sulfate crystals, and 1.2 for oleic acid droplets. Thus, all three ratios were within 0.2 of unity. The highest concentration was measured for the \SI{4.9}{\micro\meter} oleic acid droplet condition, for which the APS and the HoloCMA reported \SI{54.3}{particles\per\centi\meter\cubed} and \SI{45.9}{particles\per\centi\meter\cubed}, respectively. The PSL-sphere measurements also showed good concentration agreement over the matched \SIrange{5.0}{9.0}{\micro\meter} interval. The APS and the HoloCMA reported $1.4\times10^{-1}$ and $1.5\times10^{-1}$~\si{particles\per\centi\meter\cubed}, respectively, giving $R_n=0.9$.

At larger particle sizes, the HoloCMA generally reported higher concentrations than the APS. For sodium chloride crystals, $R_n$ decreased from 1.0 at \SI{6.1}{\micro\meter} to 0.6 at \SI{9.7}{\micro\meter}, meaning that the APS concentration was approximately 40\% lower than the HoloCMA concentration at the largest condition. For ammonium sulfate crystals, the ratio decreased from 0.9 at \SI{6.1}{\micro\meter} to 0.8 at \SI{7.7}{\micro\meter}. For oleic acid droplets, it decreased from 0.7 at \SI{6.9}{\micro\meter} to 0.5 at \SI{9.9}{\micro\meter} and remained near 0.5 at \SI{11.8}{\micro\meter}. The plateau at the largest oleic acid droplet condition indicates that the decline was not strictly monotonic within every material series, although the overall concentration ratio decreased with increasing APS aerodynamic diameter.

The size-dependent concentration difference is consistent with published size-dependent transport and counting efficiencies of the APS~\citep{volckens2005counting,vasilatou2023extending}, but the present measurements cannot attribute the trend to the APS alone. The branch-loss adjustment accounts only for predicted differential transport upstream of the two instrument inlets and does not correct for internal losses or counting efficiencies. The observed trend may also include contributions from the detection efficiency and sampled-volume calibration of the HoloCMA, flow-split effects, particle deposition, and run-to-run variability. The concentration comparison should therefore be interpreted as an inter-instrument trend rather than an absolute assessment of the accuracy of either instrument.

Overall, Figs.~\ref{fig:psl_distribution} and~\ref{fig:main_size_distributions} show that the HoloCMA reproduces the principal size populations measured by the corrected APS and tracks the expected increase in diameter across multiple particle types. Across the 12 quantitative comparisons, the mean absolute difference between the geometric means was \SI{0.5}{\micro\meter} and the mean absolute relative difference was 8.4\%, with the largest differences occurring near the lower detection boundary. Figure~\ref{fig:concentration_comparison} further shows close concentration agreement near \SI{5.0}{\micro\meter} and a decreasing APS-to-HoloCMA concentration ratio at larger sizes. Together, these results establish the geometric-sizing capability of the HoloCMA under the tested conditions and document a size-dependent inter-instrument concentration trend whose physical origin cannot be resolved from the present measurements alone.

\section{Conclusion and Discussion}
\label{sec:conclusion}

We presented the HoloCMA, a holographic instrument for real-time, in situ characterization of individual CMAs. The HoloCMA integrates a pulsed-laser digital inline holography module, active aerosol sampling, edge and GPU computing, and a deep learning-assisted processing pipeline. Acquired holograms undergo background removal, multiscale particle detection and region-of-interest extraction, numerical reconstruction and focusing, contour-based segmentation, quality control, and particle-level reporting. The current configuration records 8 MP holograms at \SI{20.0}{Hz}, supports open-path sampling and active sampling at a standard inlet flow rate of \SI{20.0}{L\per\minute}, and can operate at active inlet flow rates up to \SI{30.0}{L\per\minute}. The integrated laptop, equipped with an NVIDIA GeForce RTX 5070 Laptop GPU (\SI{8}{GB} VRAM), provides real-time computation without sustained queue buildup at concentrations up to \SI{11.4}{particles\per\centi\meter\cubed}. Additional computing capacity can extend real-time processing to the current operating limit of \SI{100.0}{particles\per\centi\meter\cubed}. The validation data were collected in active-sampling mode at an inlet flow rate of \SI{25.0}{L\per\minute}. The present optical configuration has a practical lower detection boundary of approximately \SI{5.0}{\micro\meter} and is designed to support detection up to the millimeter scale. However, the experimental validation in this study was limited to nominal diameters from \SIrange{4.9}{11.8}{\micro\meter}. Laboratory validation using PSL spheres, sodium chloride crystals, ammonium sulfate crystals, and oleic acid droplets evaluated a traceable polymer standard, crystalline particles, and liquid droplets. The normalized distributions from the HoloCMA and the corrected APS overlapped across their central size regions. Across 12 quantitative comparisons, the absolute differences between the HoloCMA and the corrected APS geometric mean diameters ranged from less than \SI{0.1}{\micro\meter} to \SI{1.0}{\micro\meter}, with mean absolute and mean absolute relative differences of \SI{0.5}{\micro\meter} and 8.4\%, respectively. Together, these results establish the individual-particle geometric-sizing capability of the HoloCMA under the tested particle and concentration conditions.

By targeting particles from approximately \SI{5.0}{\micro\meter} to the millimeter scale, the HoloCMA addresses a portion of the aerosol size spectrum that remains incompletely covered by continuous, particle-resolved measurements. Widely used instruments such as the APS and common optical particle counters typically have upper size limits near \SIrange{10.0}{20.0}{\micro\meter}, making larger dust particles, pollen, spores, and other CMAs more difficult to characterize in real time. The value of the HoloCMA, however, extends beyond its designed size range. Rather than inferring particle dimensions from aerodynamic behavior or light-scattering intensity, the HoloCMA reconstructs each particle and determines its ECD from the imaged contour. It therefore reports geometric size without material-specific conversions involving particle density, dynamic shape factor, or refractive index, which can be poorly constrained for irregular natural aerosols. Moreover, each measurement retains a reconstructed in-focus image and the associated holographic information, creating an archive of individual particles rather than only an aggregate size distribution. These records can be revisited for detailed post-acquisition examination and reprocessed as analysis methods improve. Future work can use this information to quantify particle morphology, classify known particle types, and flag particles or populations that are not represented in existing classes. Although these classification capabilities were not evaluated in the present study, they represent a natural extension of the HoloCMA framework.

It is important to note that, although the HoloCMA expands the information available for particle characterization, comparisons with established aerosol instruments must account for differences in their underlying measurement principles and reported quantities. This consideration is particularly important when comparing the HoloCMA with the APS. The APS determines aerodynamic diameter from particle time of flight, whereas the HoloCMA derives geometric diameter from reconstructed particle contours and retains images for morphological and optical analysis. These measurements are complementary: aerodynamic diameter is directly relevant to particle transport and deposition, while geometric diameter describes physical particle dimensions. Direct comparison therefore required conversion of the APS measurements to an equivalent-volume geometric diameter using assumptions about effective density, dynamic shape factor, and the behavior of oleic acid droplets within the APS. This conversion reflects the difference between the measured quantities and should not be interpreted as a limitation of the APS or an error in its aerodynamic sizing. Moreover, the FMAG diameters represent expected generation conditions rather than independent measurements of the resulting aerosols. Residual sizing differences therefore include uncertainty from aerosol generation and conversion inputs for the APS, particularly effective density and shape factor for sodium chloride crystals and ammonium sulfate crystals and the APS response to oleic acid droplets, together with morphology- and contrast-dependent focusing and contour estimation by the HoloCMA. Size-dependent discrepancies are also expected near the lower detection boundary of the HoloCMA, where pixel-limited contour resolution becomes increasingly important. Differences in binning and sampled volume further affect the comparison. Within this context, the overlap of the central size distributions and the agreement between the geometric means demonstrate consistency between two instruments operating according to fundamentally different measurement principles.

The concentration comparison likewise warrants caution. Near \SI{5.0}{\micro\meter}, the mean ratios of the APS concentration to the HoloCMA concentration were 1.1 for sodium chloride crystals, 1.0 for ammonium sulfate crystals, and 1.2 for oleic acid droplets, while the ratio generally decreased at larger sizes. The closer agreement for sodium chloride crystals and ammonium sulfate crystals than for oleic acid droplets may reflect particle-phase-dependent sampling and detection. Sodium chloride crystals, ammonium sulfate crystals, and smooth oleic acid droplets produce different holographic signatures, which can affect particle recovery near the lower detection boundary of the HoloCMA. Other inline holographic systems have similarly shown that small-droplet recovery depends strongly on particle size and position within the imaging volume~\citep{thiede2025inline}. Oleic acid droplets may also adhere more readily to sampling surfaces and are therefore less likely to remain in or re-enter the sampled flow. This contribution was not independently quantified and is expected to become more important at larger diameters. Although consistent with published size-dependent particle transport and counting efficiencies of the APS~\citep{volckens2005counting,vasilatou2023extending}, the broader concentration trend may also reflect the detection efficiency and sampled-volume calibration of the HoloCMA, flow-split effects, particle deposition, and run-to-run variability. The concentration results should therefore be interpreted as an inter-instrument trend rather than an absolute accuracy assessment of either instrument.

The present workflow is designed for operating concentrations up to \SI{100.0}{particles\per\centi\meter\cubed}. This upper bound is set by processing efficiency rather than a hard optical constraint. Because particles distributed throughout the \SI{12.0}{\milli\meter} imaging depth are projected onto the camera's \(x\)--\(y\) plane, increasing the volumetric concentration increases the number and potential overlap of particle signatures in each hologram, limiting the efficiency of the present algorithm. More advanced processing methods~\citep{shao2019hybrid,shao2020field,shao2020distribution,kumar2024sprays,shyamkumar2024generalizable} could extend this range by approximately one order of magnitude, but would increase processing time and were not used in this study.

The present results identify several priorities for extending the HoloCMA. Future validation will expand beyond the \SIrange{4.9}{11.8}{\micro\meter} range examined here to encompass the broader coarse-mode and millimeter-scale ranges. These studies will also establish a comprehensive concentration uncertainty budget using independent size and concentration references. The HoloCMA currently analyzes an optical sample volume equivalent to approximately 6.3~L/hour. This sampling throughput could be increased by approximately an order of magnitude through multipulse laser acquisition, higher-frame-rate imaging, enhanced data-acquisition hardware, and accelerated reconstruction and analysis algorithms. The modular architecture also enables the HoloCMA to be readily converted from active sampling to open-path operation by removing the active sampling module without modifying the core optical or data-processing units. Further work will develop and validate particle classification using the morphology, diffraction patterns, phase, and other optical information retained in the holographic measurements. An ongoing deployment at the ARM Bankhead National Forest observatory provides an initial field setting for this evaluation~\citep{kuang2026bnf}. Long-term, unattended field deployments will evaluate system robustness under varying aerosol concentrations, temperature, and humidity, including optical-window contamination, flow stability, and calibration drift. Together, these developments will advance the HoloCMA toward versatile, long-duration measurements of individual-particle size, concentration, morphology, optical properties, and type across atmospheric, environmental, and indoor-air applications.

\section*{CRediT authorship contribution statement}
\textbf{Nikil Krishnakumar:} Methodology, Software, Investigation, Writing -- original draft. \textbf{Ryne A. Juidici:} Investigation, Data curation, Writing -- original draft (main manuscript and supplementary information), Writing -- review \& editing. \textbf{Nicholas Bravo-Frank:} Conceptualization, Methodology, Software, Writing -- review \& editing. \textbf{Shibo Wang:} Investigation, Writing -- review \& editing. \textbf{Qisheng Ou:} Investigation, Writing -- review \& editing. \textbf{Francisco J. Romay:} Investigation, Writing -- review \& editing. \textbf{Wing Lai:} Conceptualization, Supervision, Writing -- review \& editing. \textbf{Chongai Kuang:} Writing -- review \& editing. \textbf{Naruki Hiranuma:} Writing -- review \& editing. \textbf{David Y.H. Pui:} Supervision, Writing -- review \& editing. \textbf{Jiarong Hong:} Conceptualization, Supervision, Funding acquisition, Writing -- original draft, Writing -- review \& editing.

\section*{Disclosure statement}
Nikil Krishnakumar, Nicholas Bravo-Frank, Wing Lai, and Jiarong Hong are affiliated with Particle4X, Inc., which is involved in developing the HoloCMA instrument evaluated in this study. The remaining authors report no relevant financial or non-financial competing interests.

\newpage
\section*{Funding}
This work was supported by the U.S. Department of Energy (DOE) Office of Science through a Small Business Innovation Research (SBIR) Phase I award (Award No. DE-SC0025811).

\section*{Data availability statement}
The data supporting the findings of this study are available from the corresponding author upon reasonable request.

\section*{Declaration of generative AI use}
During preparation of this manuscript, the authors used OpenAI Codex (GPT-5, accessed August 2026) to assist with manuscript organization, language refinement, \LaTeX{} formatting, and refinement of plotting-format code for the purpose of improving clarity and consistency. Generative AI was not used to generate experimental data or calculate the reported results. All analyses, reported numerical results, and scientific conclusions were produced and verified by the authors. All AI-assisted wording, formatting, and code refinements were reviewed and revised by the authors, who take full responsibility for the accuracy, originality, references, and integrity of the manuscript.

\bibliographystyle{tfcad}
\bibliography{HoloCMA_References}

\begin{thebibliography}{51}
\newcommand{\enquote}[1]{``#1''}
\providecommand{\natexlab}[1]{#1}
\providecommand{\url}[1]{\normalfont{#1}}
\providecommand{\urlprefix}{}

\bibitem[Aguilera et~al.(2025)]{aguilera2025wildfirepm}
Aguilera, Rosana, Noemie Letellier, Rupa Basu, Ambarish Vaidyanathan, Joan~A.
  Casey, Alexander Gershunov, Minghui Diao, and Tarik Benmarhnia. 2025.
  ``Effects of Multiple Wildfire Smoke Pollutants ({$\mathrm{PM}_{2.5}$},
  {$\mathrm{PM}_{10}$}, and Ozone) on Respiratory and Cardiovascular
  Hospitalizations in {California} (2006--2019).'' \emph{GeoHealth} 9 (12):
  e2025GH001510. https://doi.org/{10.1029/2025GH001510}.

\bibitem[Baldocchi et~al.(2001)]{baldocchi2001fluxnet}
Baldocchi, Dennis, Eva Falge, Lianhong Gu, Richard Olson, David Hollinger,
  Steven Running, Peter Anthoni, et~al. 2001. ``{FLUXNET}: A New Tool to Study
  the Temporal and Spatial Variability of Ecosystem Scale Carbon Dioxide, Water
  Vapor, and Energy Flux Densities.'' \emph{Bulletin of the American
  Meteorological Society} 82 (11): 2415--2434.
  https://doi.org/{10.1175/1520-0477(2001)082<2415:FANTTS>2.3.CO;2}.

\bibitem[Baron et~al.(2008)]{baron2008droplets}
Baron, Paul, Gregory~J. Deye, Anthony~B. Martinez, Erica~N. Jones, and James~S.
  Bennett. 2008. ``Size shifts in measurements of droplets with the aerodynamic
  particle sizer and the {Aerosizer}.'' \emph{Aerosol Science and Technology}
  42 (3): 201--209. https://doi.org/{10.1080/02786820801958734}.

\bibitem[Beres et~al.(2024)]{beres2024microplastics}
Beres, Nicholas~D., Julia Burkart, Elias Graf, Yanick Zeder, Lea~Ann Dailey,
  and Bernadett Weinzierl. 2024. ``Merging holography, fluorescence, and
  machine learning for in situ continuous characterization and classification
  of airborne microplastics.'' \emph{Atmospheric Measurement Techniques} 17:
  6945--6964. https://doi.org/{10.5194/amt-17-6945-2024}.

\bibitem[Berg(2022)]{berg2022tutorial}
Berg, Matthew~J. 2022. ``Tutorial: {Aerosol} characterization with digital
  in-line holography.'' \emph{Journal of Aerosol Science} 165: 106023.
  https://doi.org/{10.1016/j.jaerosci.2022.106023}.

\bibitem[Boucher et~al.(2013)]{boucher2013clouds}
Boucher, Olivier, David Randall, Paulo Artaxo, Christopher Bretherton, Graham
  Feingold, Piers Forster, Veli-Matti Kerminen, et~al. 2013. ``Clouds and
  Aerosols.'' In \emph{Climate Change 2013: The Physical Science Basis.
  Contribution of Working Group I to the Fifth Assessment Report of the
  Intergovernmental Panel on Climate Change},  edited by Thomas~F. Stocker,
  Dahe Qin, Gian-Kasper Plattner, Melinda Tignor, Simon~K. Allen, Judith
  Boschung, Alexander Nauels, Yu~Xia, Vincent Bex, and Pauline~M. Midgley,
  571--657. Cambridge, UK and New York, NY: Cambridge University Press.

\bibitem[Bravo-Frank, Feng, and Hong(2024)]{bravo2024ham}
Bravo-Frank, Nicholas, Lei Feng, and Jiarong Hong. 2024. ``Holographic
  Air-Quality Monitor ({HAM}).'' \emph{Indoor Air} 2024: 2210837.
  https://doi.org/{10.1155/2024/2210837}.

\bibitem[Bravo-Frank et~al.(2024)]{bravofrank2024bacteria}
Bravo-Frank, Nicholas, Rushikesh Zende, Lei Feng, Nicolas Mesyngier, Aditya
  Pachpute, and Jiarong Hong. 2024. ``Realtime bacteria detection and analysis
  in sterile liquid products using deep learning holographic imaging.''
  \emph{npj Biosensing} 1 (1): 8. https://doi.org/{10.1038/s44328-024-00008-9}.

\bibitem[D'Amato et~al.(2010)]{damato2010effects}
D'Amato, Gennaro, Lorenzo Cecchi, Maria D'Amato, and Gennaro Liccardi. 2010.
  ``Urban air pollution and climate change as environmental risk factors of
  respiratory allergy: an update.'' \emph{Journal of Investigational
  Allergology and Clinical Immunology} 20 (2): 95--102.

\bibitem[Douwes et~al.(2003)]{douwes2003bioaerosol}
Douwes, Jeroen, Peter Thorne, Neil Pearce, and Dick Heederik. 2003.
  ``Bioaerosol Health Effects and Exposure Assessment: Progress and
  Prospects.'' \emph{The Annals of Occupational Hygiene} 47 (3): 187--200.
  https://doi.org/{10.1093/annhyg/meg032}.

\bibitem[Duan et~al.(2016)]{duan2016generation}
Duan, Hongxu, Francisco~J. Romay, Cheng Li, Amir Naqwi, Weiwei Deng, and
  Benjamin Y.~H. Liu. 2016. ``Generation of monodisperse aerosols by combining
  aerodynamic flow-focusing and mechanical perturbation.'' \emph{Aerosol
  Science and Technology} 50 (1): 17--25.
  https://doi.org/{10.1080/02786826.2015.1123213}.

\bibitem[Fugal and Shaw(2009)]{fugal2009cloud}
Fugal, Jacob~P., and Raymond~A. Shaw. 2009. ``Cloud particle size distributions
  measured with an airborne digital in-line holographic instrument.''
  \emph{Atmospheric Measurement Techniques} 2 (1): 259--271.
  https://doi.org/{10.5194/amt-2-259-2009}.

\bibitem[Hinds and Zhu(2022)]{Hinds2022Aerosol}
Hinds, William~C., and Yifang Zhu. 2022. \emph{Aerosol Technology: Properties,
  Behavior, and Measurement of Airborne Particles}. 3rd ed. Hoboken, NJ: John
  Wiley \& Sons.

\bibitem[Howard et~al.(2019)]{howard2019mobilenetv3}
Howard, Andrew, Mark Sandler, Grace Chu, Liang-Chieh Chen, Bo~Chen, Mingxing
  Tan, Weijun Wang, et~al. 2019. ``Searching for {MobileNetV3}.'' In
  \emph{Proceedings of the IEEE/CVF International Conference on Computer
  Vision}, 1314--1324.

\bibitem[{International Organization for
  Standardization}(2015)]{iso14644cleanrooms}
{International Organization for Standardization}. 2015. ``{ISO} 14644-1:2015
  Cleanrooms and Associated Controlled Environments -- Part 1: Classification
  of Air Cleanliness by Particle Concentration.'' .

\bibitem[Katz and Sheng(2010)]{katz2010applications}
Katz, Joseph, and Jian Sheng. 2010. ``Applications of holography in fluid
  mechanics and particle dynamics.'' \emph{Annual Review of Fluid Mechanics}
  42: 531--555. https://doi.org/{10.1146/annurev-fluid-121108-145508}.

\bibitem[Kiselev, Bonacina, and Wolf(2013)]{kiselev2013pollen}
Kiselev, Denis, Luigi Bonacina, and Jean-Pierre Wolf. 2013. ``A flash-lamp
  based device for fluorescence detection and identification of individual
  pollen grains.'' \emph{Review of Scientific Instruments} 84 (3): 033302.
  https://doi.org/{10.1063/1.4793792}.

\bibitem[Kok et~al.(2017)]{kok2017dust}
Kok, Jasper~F., David~A. Ridley, Qing Zhou, Ron~L. Miller, Chun Zhao,
  Colette~L. Heald, Daniel~S. Ward, Samuel Albani, and Karsten Haustein. 2017.
  ``Smaller desert dust cooling effect estimated from analysis of dust size and
  abundance.'' \emph{Nature Geoscience} 10: 274--278.
  https://doi.org/{10.1038/ngeo2912}.

\bibitem[Kuang et~al.(2026)]{kuang2026bnf}
Kuang, Chongai, Scott~E. Giangrande, Shawn~P. Serbin, Patty Campbell,
  Gregory~S. Elsaesser, Pierre Gentine, Thijs Heus, et~al. 2026. ``The {U.S.
  DOE ARM} User Facility Establishes a New Site for Studies of
  Land--Aerosol--Cloud Interactions in the {Southeastern United States}.''
  \emph{Bulletin of the American Meteorological Society} 107: E1--E8.
  https://doi.org/{10.1175/BAMS-D-25-0072.1}.

\bibitem[Kulkarni, Baron, and Willeke(2011)]{baron2011aerosol}
Kulkarni, Pramod, Paul~A. Baron, and Klaus Willeke, eds. 2011. \emph{Aerosol
  Measurement: Principles, Techniques, and Applications}. 3rd ed. Hoboken, NJ:
  John Wiley \& Sons.

\bibitem[Kuwata and Kondo(2009)]{Kuwata2009}
Kuwata, Mikinori, and Yutaka Kondo. 2009. ``Measurements of particle masses of
  inorganic salt particles for calibration of cloud condensation nuclei
  counters.'' \emph{Atmospheric Chemistry and Physics} 9 (16): 5921--5932.
  https://doi.org/{10.5194/acp-9-5921-2009}.

\bibitem[Lieberherr et~al.(2021)]{lieberherr2021assessment}
Lieberherr, Gian, Kevin Auderset, Bertrand Calpini, Bernard Clot, Beno{\^i}t
  Crouzy, Martin Gysel-Beer, Thomas Konzelmann, et~al. 2021. ``Assessment of
  real-time bioaerosol particle counters using reference chamber experiments.''
  \emph{Atmospheric Measurement Techniques} 14 (12): 7693--7706.
  https://doi.org/{10.5194/amt-14-7693-2021}.

\bibitem[Lin et~al.(2017)]{lin2017fpn}
Lin, Tsung-Yi, Piotr Doll{\'a}r, Ross Girshick, Kaiming He, Bharath Hariharan,
  and Serge Belongie. 2017. ``Feature Pyramid Networks for Object Detection.''
  In \emph{Proceedings of the IEEE Conference on Computer Vision and Pattern
  Recognition}, 936--944.

\bibitem[Mather and Voyles(2013)]{mather2013arm}
Mather, J.~H., and J.~W. Voyles. 2013. ``The {ARM} Climate Research Facility: A
  Review of Structure and Capabilities.'' \emph{Bulletin of the American
  Meteorological Society} 94 (3): 377--392.
  https://doi.org/{10.1175/BAMS-D-11-00218.1}.

\bibitem[Matsushima and Shimobaba(2009)]{matsushima2009bandlimited}
Matsushima, Kyoji, and Tomoyoshi Shimobaba. 2009. ``Band-Limited Angular
  Spectrum Method for Numerical Simulation of Free-Space Propagation in Far and
  Near Fields.'' \emph{Optics Express} 17 (22): 19662--19673.
  https://doi.org/{10.1364/OE.17.019662}.

\bibitem[McCartney(1994)]{mccartney1994spores}
McCartney, H.~Alastair. 1994. ``Dispersal of Spores and Pollen from Crops.''
  \emph{Grana} 33 (2): 76--80. https://doi.org/{10.1080/00173139409427835}.

\bibitem[McMurry(2000)]{mcmurry2000review}
McMurry, Peter~H. 2000. ``A review of atmospheric aerosol measurements.''
  \emph{Atmospheric Environment} 34 (12--14): 1959--1999.
  https://doi.org/{10.1016/S1352-2310(99)00455-0}.

\bibitem[Meng et~al.(2024)]{meng2024pharmaceutical}
Meng, Han, Angus Shiue, Chenhua Wang, Junjie Liu, Lizhi Jia, and Graham
  Leggett. 2024. ``Particle and bacterial colony emissions from garments and
  humans in pharmaceutical cleanrooms.'' \emph{Journal of Building Engineering}
  97: 110829. https://doi.org/{10.1016/j.jobe.2024.110829}.

\bibitem[Myhre et~al.(2013)]{myhre2013anthropogenic}
Myhre, Gunnar, Drew Shindell, Fran{\c{c}}ois-Marie Br{\'e}on, William Collins,
  Jan Fuglestvedt, Jianping Huang, Dorothy Koch, et~al. 2013. ``Anthropogenic
  and Natural Radiative Forcing.'' In \emph{Climate Change 2013: The Physical
  Science Basis. Contribution of Working Group I to the Fifth Assessment Report
  of the Intergovernmental Panel on Climate Change},  edited by Thomas~F.
  Stocker, Dahe Qin, Gian-Kasper Plattner, Melinda Tignor, Simon~K. Allen,
  Judith Boschung, Alexander Nauels, Yu~Xia, Vincent Bex, and Pauline~M.
  Midgley, 659--740. Cambridge, UK and New York, NY: Cambridge University
  Press.

\bibitem[{NVIDIA Corporation}(2023)]{nvidia2023tensorrt}
{NVIDIA Corporation}. 2023. \emph{{NVIDIA TensorRT Developer Guide}}. NVIDIA
  Corporation. Version 8.6.1,
  \urlprefix\url{https://docs.nvidia.com/deeplearning/tensorrt/archives/tensorrt-861/developer-guide/index.html}.

\bibitem[Paszke et~al.(2019)]{paszke2019pytorch}
Paszke, Adam, Sam Gross, Francisco Massa, Adam Lerer, James Bradbury, Gregory
  Chanan, Trevor Killeen, et~al. 2019. ``{PyTorch}: An Imperative Style,
  High-Performance Deep Learning Library.'' In \emph{Advances in Neural
  Information Processing Systems}, Vol.~32, 8024--8035.

\bibitem[Pich(1972)]{pich1972theory}
Pich, Josef. 1972. ``Theory of gravitational deposition of particles from
  laminar flows in channels.'' \emph{Aerosol Science} 3: 351--361.
  https://doi.org/{10.1016/0021-8502(72)90090-0}.

\bibitem[Reid et~al.(2016)]{reid2016wildfire}
Reid, Colleen~E., Michael Brauer, Fay~H. Johnston, Michael Jerrett, John~R.
  Balmes, and Catherine~T. Elliott. 2016. ``Critical Review of Health Impacts
  of Wildfire Smoke Exposure.'' \emph{Environmental Health Perspectives} 124
  (9): 1334--1343. https://doi.org/{10.1289/ehp.1409277}.

\bibitem[Sauvageat et~al.(2020)]{sauvageat2020pollen}
Sauvageat, Eric, Yanick Zeder, Kevin Auderset, Bertrand Calpini, Bernard Clot,
  Beno{\^i}t Crouzy, Thomas Konzelmann, Gian Lieberherr, Fiona Tummon, and
  Konstantina Vasilatou. 2020. ``Real-time pollen monitoring using digital
  holography.'' \emph{Atmospheric Measurement Techniques} 13 (3): 1539--1550.
  https://doi.org/{10.5194/amt-13-1539-2020}.

\bibitem[Savage et~al.(2017)]{savage2017wibs}
Savage, Nicole~J., Christine~E. Krentz, Tobias K{\"o}nemann, Taewon~T. Han,
  Gediminas Mainelis, Christopher P{\"o}hlker, and J.~Alex Huffman. 2017.
  ``Systematic characterization and fluorescence threshold strategies for the
  wideband integrated bioaerosol sensor ({WIBS}) using size-resolved biological
  and interfering particles.'' \emph{Atmospheric Measurement Techniques} 10
  (11): 4279--4302. https://doi.org/{10.5194/amt-10-4279-2017}.

\bibitem[Seinfeld and Pandis(2016)]{seinfeld2016atmospheric}
Seinfeld, John~H., and Spyros~N. Pandis. 2016. \emph{Atmospheric Chemistry and
  Physics: From Air Pollution to Climate Change}. 3rd ed. Hoboken, NJ: John
  Wiley \& Sons.

\bibitem[Shao, Li, and Hong(2019)]{shao2019hybrid}
Shao, Siyao, Cheng Li, and Jiarong Hong. 2019. ``A hybrid image processing
  method for measuring {3D} bubble distribution using digital inline
  holography.'' \emph{Chemical Engineering Science} 207: 929--941.
  https://doi.org/{10.1016/j.ces.2019.07.009}.

\bibitem[Shao, Mallery, and Hong(2020)]{shao2020distribution}
Shao, Siyao, Kevin Mallery, and Jiarong Hong. 2020. ``Machine learning
  holography for measuring {3D} particle distribution.'' \emph{Chemical
  Engineering Science} 225: 115830.
  https://doi.org/{10.1016/j.ces.2020.115830}.

\bibitem[Shao et~al.(2020)]{shao2020field}
Shao, Siyao, Kevin Mallery, S.~Santosh Kumar, and Jiarong Hong. 2020. ``Machine
  learning holography for {3D} particle field imaging.'' \emph{Optics Express}
  28 (3): 2987--2999. https://doi.org/{10.1364/OE.379480}.

\bibitem[Shyam~Kumar et~al.(2024)]{kumar2024sprays}
Shyam~Kumar, M., Christopher~J. Hogan, Steven~A. Fredericks, and Jiarong Hong.
  2024. ``Visualization and characterization of agricultural sprays using
  machine learning based digital inline holography.'' \emph{Computers and
  Electronics in Agriculture} 216: 108486.
  https://doi.org/{10.1016/j.compag.2023.108486}.

\bibitem[Shyam~Kumar and Hong(2025)]{shyamkumar2025review}
Shyam~Kumar, M., and Jiarong Hong. 2025. ``A review of {3D} particle tracking
  and flow diagnostics using digital holography.'' \emph{Measurement Science
  and Technology} 36 (3): 032005. https://doi.org/{10.1088/1361-6501/adabff}.

\bibitem[{Shyam Kumar M.} and Hong(2024)]{shyamkumar2024generalizable}
{Shyam Kumar M.}, and Jiarong Hong. 2024. ``Generalizable deep learning
  approach for {3D} particle imaging using holographic microscopy ({HM}).''
  \emph{Optics Express} 32 (27): 48159--48173.
  https://doi.org/{10.1364/OE.535207}.

\bibitem[Sulvar{\'a}n-Salmoreno, Moreno-Hern{\'a}ndez, and
  Torres-Armenta(2026)]{sulvaran2026benchmarking}
Sulvar{\'a}n-Salmoreno, Brandon~R., David Moreno-Hern{\'a}ndez, and Diego
  Torres-Armenta. 2026. ``Benchmarking Focus Metrics for Microparticle
  Localization in In-Line Digital Holography.'' \emph{Optics} 7 (3): 38.
  https://doi.org/{10.3390/opt7030038}.

\bibitem[Suzuki and Abe(1985)]{suzuki1985border}
Suzuki, Satoshi, and Keiichi Abe. 1985. ``Topological structural analysis of
  digitized binary images by border following.'' \emph{Computer Vision,
  Graphics, and Image Processing} 30 (1): 32--46.
  https://doi.org/{10.1016/0734-189X(85)90016-7}.

\bibitem[Thiede et~al.(2025)]{thiede2025inline}
Thiede, Birte, Oliver Schlenczek, Katja Stieger, Alexander Ecker, Eberhard
  Bodenschatz, and Gholamhossein Bagheri. 2025. ``In-line holographic droplet
  imaging: accelerated classification with convolutional neural networks and
  quantitative experimental validation.'' \emph{Atmospheric Measurement
  Techniques} 18: 6291--6314. https://doi.org/{10.5194/amt-18-6291-2025}.

\bibitem[Tiwari, Li, and Romay(2017)]{tiwari2017fmag}
Tiwari, Andrea~J., Lin Li, and Francisco~J. Romay. 2017. ``Influence of shape
  factor and effective density on aerodynamic sizing of particles generated by
  the {MSP/TSI} 1520 flow-focusing monodisperse aerosol generator ({FMAG}).''
  In \emph{American Association for Aerosol Research Annual Conference},
  Raleigh, NC.

\bibitem[{TSI Incorporated}(2013)]{tsi_aps3321}
{TSI Incorporated}. 2013. \emph{Aerodynamic Particle Sizer ({APS}) Spectrometer
  Model 3321 Operation and Service Manual}. Shoreview, MN: TSI Incorporated.
  P/N 1980571, Revision G.

\bibitem[Van~der Heyden et~al.(2021)]{vanderheyden2021monitoring}
Van~der Heyden, Herv{\'e}, Pierre Dutilleul, Jean-Benoit Charron, Guillaume~J.
  Bilodeau, and Odile Carisse. 2021. ``Monitoring airborne inoculum for
  improved plant disease management. A review.'' \emph{Agronomy for Sustainable
  Development} 41: 40. https://doi.org/{10.1007/s13593-021-00694-z}.

\bibitem[Vasilatou et~al.(2023)]{vasilatou2023extending}
Vasilatou, Konstantina, Christian W{\"a}lchli, Kenjiro Iida, Stefan Horender,
  Torsten Tritscher, Tobias Hammer, Jenny Rissler, Fran{\c{c}}ois Gaie-Levrel,
  and Kevin Auderset. 2023. ``Extending traceability in airborne particle size
  distribution measurements beyond 10~{\textmu}m: counting efficiency and
  unit-to-unit variability of four aerodynamic particle size spectrometers.''
  \emph{Aerosol Science and Technology} 57 (1): 24--34.
  https://doi.org/{10.1080/02786826.2022.2139659}.

\bibitem[Volckens and Peters(2005)]{volckens2005counting}
Volckens, John, and Thomas~M. Peters. 2005. ``Counting and particle
  transmission efficiency of the aerodynamic particle sizer.'' \emph{Journal of
  Aerosol Science} 36: 1400--1408.
  https://doi.org/{10.1016/j.jaerosci.2005.03.009}.

\bibitem[Zelenyuk, Cai, and Imre(2006)]{zelenyuk2006agglomerates}
Zelenyuk, Alla, Yong Cai, and Dan Imre. 2006. ``From agglomerates of spheres to
  irregularly shaped particles: determination of dynamic shape factors from
  measurements of mobility and vacuum aerodynamic diameters.'' \emph{Aerosol
  Science and Technology} 40 (3): 197--217.
  https://doi.org/{10.1080/02786820500529406}.

\end{thebibliography}

\clearpage
\begingroup
\pdfpagewidth=8.5in
\pdfpageheight=11in
\hoffset=-1in
\voffset=-1in
\newcommand{\appendsupplementpage}[1]{%
  \pdfximage page #1 {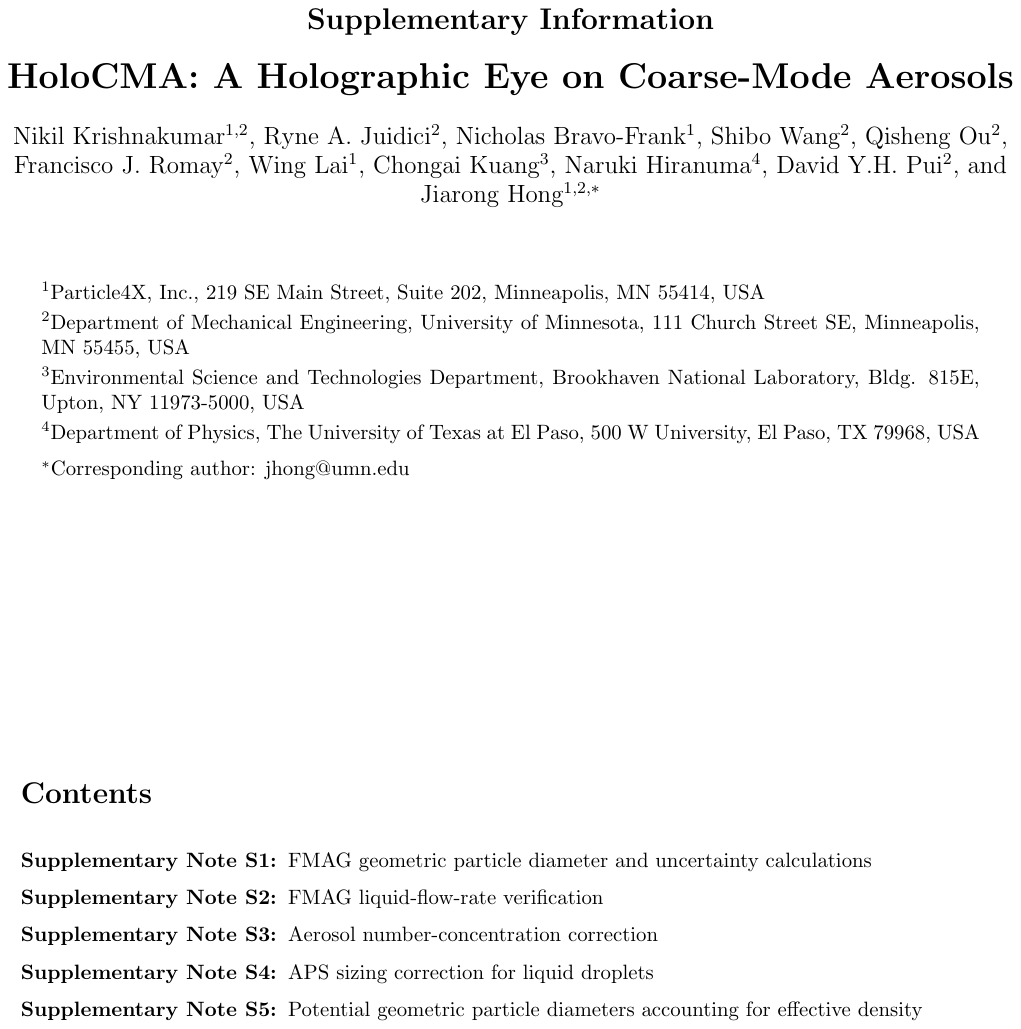}%
  \shipout\vbox to \pdfpageheight{%
    \offinterlineskip
    \hbox to \pdfpagewidth{\pdfrefximage\pdflastximage\hss}%
    \vss
  }%
}
\appendsupplementpage{1}
\appendsupplementpage{2}
\appendsupplementpage{3}
\appendsupplementpage{4}
\appendsupplementpage{5}
\appendsupplementpage{6}
\appendsupplementpage{7}
\endgroup

\end{document}


\hypersetup{pageanchor=false}
\thispagestyle{empty}
\begin{center}
  \vspace*{0.45in}
  {\Large\bfseries Supplementary Information\par}
  \vspace{0.6em}
  {\LARGE\bfseries HoloCMA: A Holographic Eye on Coarse-Mode Aerosols\par}
  \vspace{1.2em}
  {\large
  Nikil Krishnakumar$^{1,2}$, Ryne A. Juidici$^{2}$, Nicholas Bravo-Frank$^{1}$, Shibo Wang$^{2}$, Qisheng Ou$^{2}$,\\
  Francisco J. Romay$^{2}$, Wing Lai$^{1}$, Chongai Kuang$^{3}$, Naruki Hiranuma$^{4}$, David Y.H. Pui$^{2}$, and Jiarong Hong$^{1,2,*}$\par}
\end{center}

\vspace{1.0em}
\begin{center}
\begin{minipage}{0.92\textwidth}
\small
$^{1}$Particle4X, Inc., 219 SE Main Street, Suite 202, Minneapolis, MN 55414, USA\\[0.25em]
$^{2}$Department of Mechanical Engineering, University of Minnesota, 111 Church Street SE, Minneapolis, MN 55455, USA\\[0.25em]
$^{3}$Environmental Science and Technologies Department, Brookhaven National Laboratory, Bldg. 815E, Upton, NY 11973-5000, USA\\[0.25em]
$^{4}$Department of Physics, The University of Texas at El Paso, 500 W University, El Paso, TX 79968, USA\\[0.55em]
$^{*}$Corresponding author: \href{mailto:jhong@umn.edu}{jhong@umn.edu}
\end{minipage}
\end{center}

\vfill
\begin{center}
\begin{minipage}{0.96\textwidth}
{\Large\bfseries Contents\par}
\vspace{0.9em}
\small
\begin{description}
  \item[Supplementary Note S1:] FMAG geometric particle diameter and uncertainty calculations
  \item[Supplementary Note S2:] FMAG liquid-flow-rate verification
  \item[Supplementary Note S3:] Aerosol number-concentration correction
  \item[Supplementary Note S4:] APS sizing correction for liquid droplets
  \item[Supplementary Note S5:] Potential geometric particle diameters accounting for effective density
\end{description}
\end{minipage}
\end{center}
\vfill

\clearpage
\hypersetup{pageanchor=true}
\vspace*{0.15in}
\suppnote{S1}{FMAG geometric particle diameter and uncertainty calculations}
As described in the manuscript, the droplet diameter produced by the flow-focusing monodisperse aerosol generator (FMAG, Model 1520) is controlled by the liquid flow rate and droplet-generation frequency~\citep{Duan2016FMAG}:
\begin{equation}
  d_d = \left(\frac{6Q}{\pi f}\right)^{1/3}.
\end{equation}
Here, $d_d$ is the droplet diameter, $Q$ is the liquid flow rate, and $f$ is the droplet-generation frequency. For a solution volume fraction $C$, the nominal geometric particle diameter after solvent evaporation is~\citep{Duan2016FMAG}
\begin{equation}
  d_p = C^{1/3}d_d.
\end{equation}
\par\smallskip
Table~S1 lists the generation parameters and propagated uncertainty for each nominal geometric particle diameter.
\begin{table}[H]
\centering
\caption{Flow-Focusing Monodisperse Aerosol Generator (FMAG, Model 1520) generation parameters for the HoloCMA evaluation.}
\small
\setlength{\tabcolsep}{4.3pt}
\renewcommand{\arraystretch}{1.12}
\begin{tabular}{@{}lccccc@{}}
\toprule
Input solution & \shortstack{Solution\\concentration\\(\si{\liter\per\liter})} & \shortstack{Frequency\\(\si{\kilo\hertz})} & \shortstack{Flow rate\\(\si{\milli\liter\per\hour})} & \shortstack{Droplet diameter\\(\si{\micro\meter})} & \shortstack{Geometric particle\\diameter (\si{\micro\meter})} \\
\midrule
Sodium chloride & \num{0.0013 \pm 0.0002} & 90.0 & 3.87 & \num{28.4 \pm 0.3} & \num{3.05 \pm 0.04} \\
Sodium chloride & \num{0.0050 \pm 0.0002} & 90.0 & 3.87 & \num{28.4 \pm 0.3} & \num{4.84 \pm 0.05} \\
Sodium chloride & \num{0.0100 \pm 0.0002} & 90.0 & 3.87 & \num{28.4 \pm 0.3} & \num{6.11 \pm 0.07} \\
Sodium chloride & \num{0.0150 \pm 0.0002} & 90.0 & 3.87 & \num{28.4 \pm 0.3} & \num{6.99 \pm 0.08} \\
Sodium chloride & \num{0.0400 \pm 0.0002} & 90.0 & 3.87 & \num{28.4 \pm 0.3} & \num{9.70 \pm 0.10} \\
\addlinespace[2pt]
Ammonium sulfate & \num{0.0050 \pm 0.0002} & 90.0 & 3.87 & \num{28.4 \pm 0.3} & \num{4.85 \pm 0.05} \\
Ammonium sulfate & \num{0.0100 \pm 0.0002} & 90.0 & 3.87 & \num{28.4 \pm 0.3} & \num{6.11 \pm 0.07} \\
Ammonium sulfate & \num{0.0200 \pm 0.0002} & 90.0 & 3.87 & \num{28.4 \pm 0.3} & \num{7.70 \pm 0.08} \\
\addlinespace[2pt]
Oleic acid & \num{0.0040 \pm 0.0004} & 86.0 & 4.83 & \num{31.0 \pm 0.3} & \num{4.92 \pm 0.06} \\
Oleic acid & \num{0.0040 \pm 0.0004} & 50.0 & 7.73 & \num{43.5 \pm 0.4} & \num{6.90 \pm 0.08} \\
Oleic acid & \num{0.0200 \pm 0.0020} & 64.0 & 5.80 & \num{36.4 \pm 0.4} & \num{9.87 \pm 0.14} \\
Oleic acid & \num{0.0200 \pm 0.0020} & 50.0 & 7.73 & \num{43.5 \pm 0.4} & \num{11.80 \pm 0.16} \\
\bottomrule
\end{tabular}
\end{table}
Uncertainties in the solution concentration, droplet diameter, and geometric particle diameter were propagated as described below.
\subsection*{Uncertainty in solution concentration}
The calculation for the uncertainty in the solution concentration was dependent on the exact solution. All of the sodium chloride solutions shared the same formula for the uncertainty in the concentration. For ammonium sulfate, the formula for the uncertainty in the concentration was dependent on the concentration of the solution. Similarly, the formula for the uncertainty in the oleic acid solutions was also dependent on the solution concentration.
\subsubsection*{Sodium chloride solutions and the 0.01 and \SI{0.02}{\liter\per\liter} ammonium sulfate solutions}
The volume fraction of each solution was calculated from
\begin{equation}
  C = \frac{m}{V_{\mathrm{tot}}\rho_{\mathrm{bulk}}},
\end{equation}
where $C$ is the solution volume fraction used in Equation~(S2), $m$ is the solute mass, $V_{\mathrm{tot}}$ is the total solution volume, and $\rho_{\mathrm{bulk}}$ is the assumed bulk density of the solute. The balance had a listed accuracy of \SI{0.02}{\gram} and a resolution of \SI{0.01}{\gram}, giving a combined mass uncertainty of \SI{0.0224}{\gram}. Solution volume was set with a \SI{100}{\milli\liter} volumetric flask. The flask uncertainty of \SI{0.1}{\milli\liter} was doubled for conservative propagation, while the bulk densities were treated as exact. The concentration uncertainty was
\begin{equation}
  U_C = \sqrt{\left(\frac{U_m}{V_{\mathrm{tot}}\rho_{\mathrm{bulk}}}\right)^2
  + \left(\frac{-mU_{V_{\mathrm{tot}}}}{V_{\mathrm{tot}}^2\rho_{\mathrm{bulk}}}\right)^2}.
\end{equation}
\subsubsection*{\SI{0.005}{\liter\per\liter} ammonium sulfate solution}
The \SI{0.005}{\liter\per\liter} ammonium sulfate solution was prepared by diluting the \SI{0.01}{\liter\per\liter} stock solution:
\begin{equation}
  C = \frac{C_{0.01}V_{0.01}}{V_{\mathrm{H_2O}}+V_{0.01}}.
\end{equation}
Here, $C_{0.01}$ and $V_{0.01}$ are the concentration and transferred volume of the stock solution, and $V_{\mathrm{H_2O}}$ is the volume of dilution water. Each liquid was transferred with a \SI{25}{\milli\liter} pipette having a listed uncertainty of \SI{0.1}{\milli\liter}. This value was doubled for conservative propagation. The diluted-solution uncertainty was
 \begin{equation}
  U_C = \sqrt{\left(\frac{U_{C_{0.01}}V_{0.01}}{V_{\mathrm{H_2O}}+V_{0.01}}\right)^2
  + \left(\frac{U_{V_{0.01}}C_{0.01}V_{\mathrm{H_2O}}}{(V_{\mathrm{H_2O}}+V_{0.01})^2}\right)^2
  + \left(\frac{-U_{V_{\mathrm{H_2O}}}C_{0.01}V_{0.01}}{(V_{\mathrm{H_2O}}+V_{0.01})^2}\right)^2}.
\end{equation}
\subsubsection*{\SI{0.02}{\liter\per\liter} oleic acid solution}
The \SI{0.02}{\liter\per\liter} oleic acid solution was prepared directly, with volume fraction
\begin{equation}
  C = \frac{V_{\mathrm{oleic}}}{V_{\mathrm{oleic}}+V_{\mathrm{ethanol}}}.
\end{equation}
$V_{\mathrm{oleic}}$ and $V_{\mathrm{ethanol}}$ are the added volumes of oleic acid and ethanol. The listed \SI{0.1}{\milli\liter} uncertainty for each volume was doubled for conservative propagation, giving
\begin{equation}
  U_C = \sqrt{\left(\frac{V_{\mathrm{ethanol}}U_{V_{\mathrm{oleic}}}}{(V_{\mathrm{oleic}}+V_{\mathrm{ethanol}})^2}\right)^2
  + \left(\frac{-V_{\mathrm{oleic}}U_{V_{\mathrm{ethanol}}}}{(V_{\mathrm{oleic}}+V_{\mathrm{ethanol}})^2}\right)^2}.
\end{equation}
\subsubsection*{\SI{0.004}{\liter\per\liter} oleic acid solution}
The \SI{0.004}{\liter\per\liter} oleic acid solution was prepared by diluting the \SI{0.02}{\liter\per\liter} stock solution:
\begin{equation}
  C = \frac{C_{0.02}V_{0.02}}{V_{\mathrm{tot}}}.
\end{equation}
Here, $C_{0.02}$ and $V_{0.02}$ are the concentration and transferred volume of the stock solution, and $V_{\mathrm{tot}}$ is the final diluted volume. A \SI{10}{\milli\liter} pipette with a listed uncertainty of \SI{0.1}{\milli\liter} was used twice to transfer \SI{20}{\milli\liter}. The combined transferred-volume uncertainty was conservatively taken as \SI{0.3}{\milli\liter}. The final volume was set with a \SI{100}{\milli\liter} volumetric flask having an assumed uncertainty of \SI{0.2}{\milli\liter}. The resulting concentration uncertainty was
\begin{equation}
  U_C = \sqrt{\left(\frac{V_{0.02}U_{C_{0.02}}}{V_{\mathrm{tot}}}\right)^2
  + \left(\frac{C_{0.02}U_{V_{0.02}}}{V_{\mathrm{tot}}}\right)^2
  + \left(\frac{-V_{0.02}C_{0.02}U_{V_{\mathrm{tot}}}}{V_{\mathrm{tot}}^2}\right)^2}.
\end{equation}
\subsection*{Uncertainty in droplet diameter}
The droplet-diameter uncertainty was taken as $\pm1\%$ of the calculated diameter, consistent with the FMAG specification and its stated liquid-flow and generation-frequency uncertainties~\citep{TSI1520Manual}.
\subsection*{Uncertainty in particle diameter}
Propagating the solution-concentration and droplet-diameter uncertainties through Equation~(S2) gives
\begin{equation}
  U_{d_p} = \sqrt{\left(U_{d_d}C^{1/3}\right)^2
  + \left(\frac{d_dU_C}{3C^{2/3}}\right)^2}.
\end{equation}
This expression was evaluated for every generated particle condition in Table~S1.
\suppnote{S2}{FMAG liquid-flow-rate verification}
Liquid flow rate is a key FMAG operating parameter and is controlled by a syringe-pump setpoint on the FMAG control panel. The syringe pump used in this study had been replaced with an updated model, so its delivered flow was verified independently. Liquid was collected in a beaker on a balance, and its accumulated mass was recorded at \SI{15}{\minute} intervals for \SI{2}{\hour}. Collected volume was calculated using an assumed liquid density of \SI{1.0}{\gram\per\centi\meter\cubed}. Figure~S1 summarizes this verification.
\begin{figure}[H]
  \centering
  \includegraphics[width=0.82\textwidth]{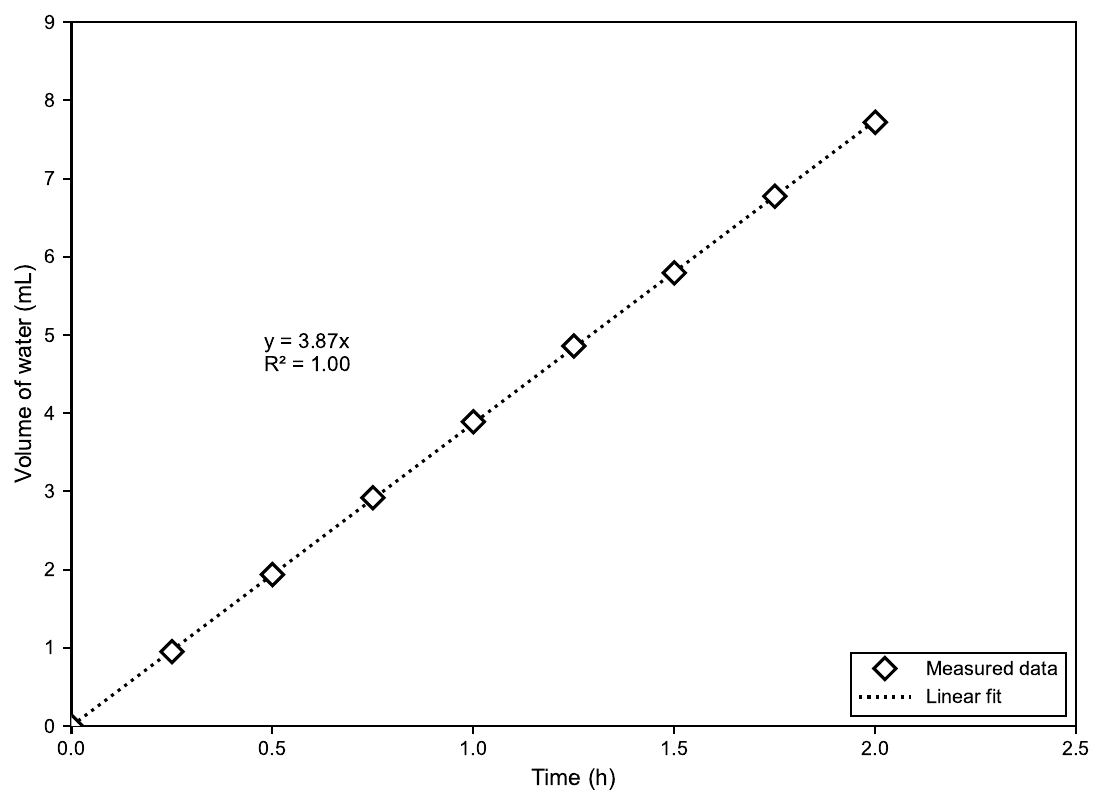}
  \caption{FMAG syringe-pump liquid-flow verification at a setpoint of \SI{4.0}{\milli\liter\per\hour}.}
\end{figure}
\begin{samepage}
At a setpoint of \SI{4.0}{\milli\liter\per\hour}, the delivered liquid flow rate was \SI{3.87}{\milli\liter\per\hour}. This measured value was used when calculating geometric particle diameter. For other setpoints, the delivered flow was estimated using
\begin{equation}
  Q = Q_{\mathrm{set}}\left(\frac{3.87}{4.00}\right),
\end{equation}
where $Q$ is the delivered liquid flow rate and $Q_{\mathrm{set}}$ is the FMAG liquid-flow setpoint.
\end{samepage}
\suppnote{S3}{Aerosol number-concentration correction}
The different inlet flow rates of the APS and the HoloCMA required different Y-junction branch diameters, producing different particle-transport losses upstream of the instrument inlets. To place both concentrations on a common upstream basis, the APS concentration was adjusted using the relative branch penetrations predicted by the gravitational-deposition model of \citet{Pich1972}. The model assumes circular sampling lines, laminar flow, and gravitational settling as the dominant loss mechanism. Table~S2 lists the branch parameters and corresponding Reynolds numbers. The gas viscosity and density were assumed to be \SI{1.8e-5}{\pascal\second} and \SI{1.2}{\kilo\gram\per\meter\cubed}, respectively. Both branches remain within the laminar-flow regime under these conditions.
 \begin{table}[H]
\centering
\caption{Key flow parameters and Reynolds numbers of the Y-junction branches.}
\small
\setlength{\tabcolsep}{8pt}
\renewcommand{\arraystretch}{1.15}
\begin{tabular}{@{}lcccc@{}}
\toprule
Branch & \shortstack{Volumetric flow rate\\(\si{\meter\cubed\per\second})} & \shortstack{Nominal diameter\\(\si{\meter})} & \shortstack{Average velocity\\(\si{\meter\per\second})} & Reynolds number \\
\midrule
APS & \num{8.3e-5} & 0.018 & 0.33 & 397 \\
HoloCMA & \num{4.17e-4} & 0.040 & 0.33 & 890 \\
\bottomrule
\end{tabular}
\end{table}
The model gives the particle penetration through a sampling line as~\citep{Pich1972}
\begin{equation}
  P = 1 - \frac{2}{\pi}\left(2\epsilon\sqrt{1-\epsilon^{2/3}}-\epsilon^{1/3}\sqrt{1-\epsilon^{2/3}} + \arcsin\!\left(\epsilon^{1/3}\right)\right),
\end{equation}
where $P$ is the penetration and $\epsilon$ is the dimensionless parameter
\begin{equation}
  \epsilon = \frac{3V_sL}{8aU}.
\end{equation}
Here, $V_s$ is the particle settling speed, $L$ is the sampling-line length, $a$ is the sampling-line radius, and $U$ is the average flow velocity. The settling speed for each particle diameter was calculated as~\citep{Hinds2022Aerosol}
\begin{equation}
  V_s = \frac{d_a^2\rho_0g}{18\mu},
\end{equation}
where $d_a$ is the aerodynamic diameter, $\rho_0$ is the aerodynamic reference density, $g$ is gravitational acceleration, and $\mu$ is the dynamic viscosity of air. Equations~(S13)--(S15) were evaluated for both Y-junction branches. Two Y-junction lengths, 6.0~in and 3.75~in, were used during the experiments, and $L$ was selected to match the junction used for each trial. Branch penetrations were calculated at every APS channel midpoint. The APS concentration was then adjusted to the equivalent concentration at the HoloCMA branch inlet using
\begin{equation}
  N_{\mathrm{APS,corr}} = N_{\mathrm{APS,raw}}\frac{P_{\mathrm{HoloCMA}}}{P_{\mathrm{APS}}}.
\end{equation}
$N_{\mathrm{APS,corr}}$ and $N_{\mathrm{APS,raw}}$ are the branch-loss-adjusted and raw APS concentrations for a given bin, respectively, and $P_{\mathrm{HoloCMA}}$ and $P_{\mathrm{APS}}$ are the predicted penetrations through the two branches at that bin midpoint. For oleic acid droplets, this concentration adjustment was applied after the liquid-droplet size-shift correction so that the corrected aerodynamic diameter was used to calculate settling velocity.

This adjustment accounts for the modeled difference in branch transport but does not correct either instrument's internal counting efficiency. The Y-junction geometry also means that the velocity profiles may remain developing over part of each branch. Based on the branch diameters and particle sizes considered here, the difference between losses under developing and fully developed flow is expected to be small, but it remains a source of model uncertainty~\citep{TAULBEE1978513}.
\suppnote{S4}{APS sizing correction for liquid droplets}
As described in the manuscript, the APS can report a size shift for large liquid droplets because droplets deform during acceleration and deposited liquid alters the velocity through the focusing nozzle. These effects shift the reported aerodynamic diameter of generated oleic acid droplets below its expected value. We applied the empirical correction of \citet{baron2008droplets}:
\begin{equation}
\Delta = \frac{-a d_a^b}{\eta^c\sigma^e}.
\end{equation} 
$\Delta$ is the diameter shift, $d_a$ is the true aerodynamic diameter, $a=\num{2.723e-4}$ is an instrument-dependent coefficient, $b=2$ is the diameter exponent, $c=0.6486$ and $e=0.3864$ are empirical exponents, $\eta$ is the liquid dynamic viscosity, and $\sigma$ is the liquid surface tension. For oleic acid, $\eta=\SI{0.0256}{\pascal\second}$ and $\sigma=\SI{0.032}{\newton\per\meter}$~\citep{baron2008droplets}. The APS-reported aerodynamic diameter is
\begin{equation}
  d_{a,\mathrm{APS}} = d_a + \Delta.
\end{equation}
Because $\Delta$ depends on $d_a$, the true aerodynamic diameter was obtained implicitly from $d_{a,\mathrm{APS}}$ using a MATLAB script. The correction was applied to every APS channel midpoint before conversion to equivalent-volume geometric diameter.
\suppnote{S5}{Potential geometric particle diameters accounting for effective density}
The nominal geometric particle diameters in Table~S1 account for uncertainty in droplet generation and solution concentration. The physical dimensions of spray-dried sodium chloride and ammonium sulfate particles can additionally differ from these nominal values because their effective densities can be lower than their bulk crystalline densities. If the density difference arises from internal voids while the particle retains an approximately spherical outer envelope, the corresponding geometric diameter can be estimated from mass conservation:
\begin{equation}
  \frac{\pi}{6}d_p^3\rho_{\mathrm{bulk}} \approx \frac{\pi}{6}d_{p,\mathrm{est}}^3\rho_e
  \quad\Longrightarrow\quad
  d_{p,\mathrm{est}} \approx \left(\frac{\rho_{\mathrm{bulk}}}{\rho_e}\right)^{1/3}d_p.
\end{equation}
Here, $d_p$ is the nominal geometric particle diameter, $\rho_{\mathrm{bulk}}$ is the bulk material density, $d_{p,\mathrm{est}}$ is the effective-density-adjusted geometric diameter, and $\rho_e$ is the effective density.

As described in the manuscript, literature-derived effective densities of \SI{1.60}{\gram\per\centi\meter\cubed} for sodium chloride and \SI{1.55}{\gram\per\centi\meter\cubed} for ammonium sulfate were used~\citep{tiwari2017fmag,Kuwata2009,zelenyuk2006agglomerates}. Table~S3 gives the corresponding diameter estimates.
\begin{table}[H]
\centering
\caption{Potential geometric particle diameters accounting for the effective density of the crystalline particles.}
\small
\setlength{\tabcolsep}{13pt}
\renewcommand{\arraystretch}{1.12}
\begin{tabular}{@{}lcc@{}}
\toprule
Input solution & \shortstack{Nominal geometric\\diameter (\si{\micro\meter})} & \shortstack{Potential geometric\\diameter (\si{\micro\meter})} \\
\midrule
Sodium chloride & \num{3.05 \pm 0.04} & \num{3.38 \pm 0.04} \\
Sodium chloride & \num{4.84 \pm 0.05} & \num{5.35 \pm 0.06} \\
Sodium chloride & \num{6.11 \pm 0.07} & \num{6.75 \pm 0.07} \\
Sodium chloride & \num{6.99 \pm 0.08} & \num{7.72 \pm 0.08} \\
Sodium chloride & \num{9.70 \pm 0.10} & \num{10.71 \pm 0.11} \\
\addlinespace[2pt]
Ammonium sulfate & \num{4.85 \pm 0.05} & \num{5.07 \pm 0.06} \\
Ammonium sulfate & \num{6.11 \pm 0.07} & \num{6.38 \pm 0.07} \\
Ammonium sulfate & \num{7.70 \pm 0.08} & \num{8.04 \pm 0.09} \\
\bottomrule
\end{tabular}
\end{table}
Equation~(S19) provides an approximate upper estimate of geometric diameter under the stated void-volume assumption. The dynamic shape factor and effective density used in the APS conversion were estimated from the literature. Because their combined ratio determines the equivalent-volume conversion, a higher effective density paired with a correspondingly higher dynamic shape factor could produce the same converted diameter. The propagated uncertainty in the estimated geometric diameter is
\begin{equation}
  U_{d_{p,\mathrm{est}}} = \left(\frac{\rho_{\mathrm{bulk}}}{\rho_e}\right)^{1/3}U_{d_p}.
\end{equation}
\par\bigskip
\small
\raggedright
\setlength{\bibsep}{0.35em}